\documentclass[12pt]{article}

\usepackage{lineno}

\usepackage[margin=1in]{geometry}
\usepackage{amssymb,amsmath,latexsym}                        
\usepackage{graphicx}
\usepackage{cite}
\usepackage{xcolor}
\usepackage{float}

\usepackage{amsthm}
\usepackage{multicol}
\usepackage{authblk}
\usepackage[colorlinks=true,linkcolor=blue, citecolor=blue, urlcolor=blue]{hyperref}

\usepackage{adjustbox}
\usepackage{bbm}
\usepackage{mathtools}

\DeclarePairedDelimiter\floor{\lfloor}{\rfloor}

\date{\today} 

\begin{document}

\title{Optimal use of Charge Information for the HL-LHC Pixel Detector Readout}
\author[1,2,3]{Yitian Chen}
\author[1,3]{Evan Frangipane}
\author[3]{Maurice Garcia-Sciveres}
\author[3]{Laura Jeanty}
\author[3]{Benjamin Nachman}
\author[3]{Simone Pagan Griso}
\author[3,4]{Fuyue Wang}
\affil[1]{\normalsize\it Department of Physics, University of California Berkeley, Berkeley, CA 94720, USA}
\affil[2]{\normalsize\it Department of Physics, Cornell University, Ithaca, NY 14853, USA}
\affil[3]{\normalsize\it Physics Division, Lawrence Berkeley National Laboratory, Berkeley, CA 94704, USA}
\affil[4]{\normalsize\it Department of Engineering Physics, Tsinghua University, Key Laboratory of Particle and Radiation Imaging, Ministry of Education, Beijing 100084, China}

\maketitle

\begin{abstract}
The pixel detectors for the High Luminosity upgrades of the ATLAS and CMS detectors will preserve digitized
charge information in spite of extremely high hit rates. Both circuit physical size and output bandwidth 
will limit the number of bits to which charge can be digitized and stored. We therefore study the 
effect of the number of bits used for digitization and storage on  single and multi-particle cluster resolution, 
efficiency, classification, and particle identification.
We show how performance degrades as fewer bits are used to digitize and to store charge. 
We find that with limited charge information (4 bits), one can achieve near optimal performance on a variety of tasks.  
\end{abstract}

\section{Introduction}
\label{sec:intro}

The experiments at the Large Hadron Collider (LHC) probe the highest energies and shortest distance scales ever measured by a terrestrial experiment. In order to further increase their sensitivity, 
the planned High Luminosity upgrades of the ATLAS~\cite{CERN-LHCC-2015-020} and CMS~\cite{Butler:2055167} pixel detectors must 
increase their readout bandwidth by a factor of 20 relative to present detectors. 
The pixel detectors provide precision three-dimensional space points for reconstruction of charged particle trajectories.
The primary position information is given by the patterns of hit pixels~\cite{binaryreadout},
but the measurement of collected electric charge in each pixel can increase the space point precision,
enable discrimination of merged hits~\cite{Aad:2014yva,Aaboud:2017all,cmsjetcore}, and be used for identification of Standard Model (SM) and 
exotic particles beyond the SM~\cite{Aaboud:2016dgf,Khachatryan:2016sfv,ATLAS:2014fka,Chatrchyan:2013oca,Aad:2011yf,Aad:2012pra,Aad:2013pqd,Chatrchyan:2012sp,Khachatryan:2011ts}.  

Analog storage and readout of charge pixel information are not feasible at high rate\footnote{During the relatively low occupancy and radiation damage environment of the LHC Run 1, CMS did read analog charge information~\cite{Karimaki:368412}.  This has been discontinued for the upgraded detector~\cite{Dominguez:1481838}.}.  
Charge information must therefore be digitized in-situ and the digital values stored on-chip for later readout.
Since the amount of on-chip storage is finite, and the number of hits per unit area that must be 
stored is determined by the hit rate and trigger latency, this results in a constraint on the number 
of charge information bits per pixel that can be stored. The RD53A chip~\cite{Garcia-Sciveres:2113263}, 
which is the prototype for both the ATLAS and CMS pixel upgrades, 
stores 4 bits per hit per 50\,$\mu$m$\times$50\,$\mu$m pixel at 75\,KHz/pixel input hit rate
and 12\,$\mu$s latency. This was a value imposed by physical space for circuitry rather than physics performance 
considerations. Instead, in this paper we investigate what is the number of bits one would need to store
in order to achieve a certain performance. In the RD53A chip the digitization of charge uses a 4-bit
Time-over-Threshold (ToT) linear analog-to-digital conversion (ADC) method (see Fig.~\ref{fig:ToTschematic}), which seems reasonable given that 4 bits are stored. However,
the number of ADC and stored bits do not have to match, because stored bits can be compressed. We therefore 
investigate the question of number of ADC bits separately from storage.  This is a generic study using standalone simulation code, and does not use ATLAS or CMS-specific simulation.

\begin{figure}[h!]
\centering
\includegraphics[width=0.55\textwidth]{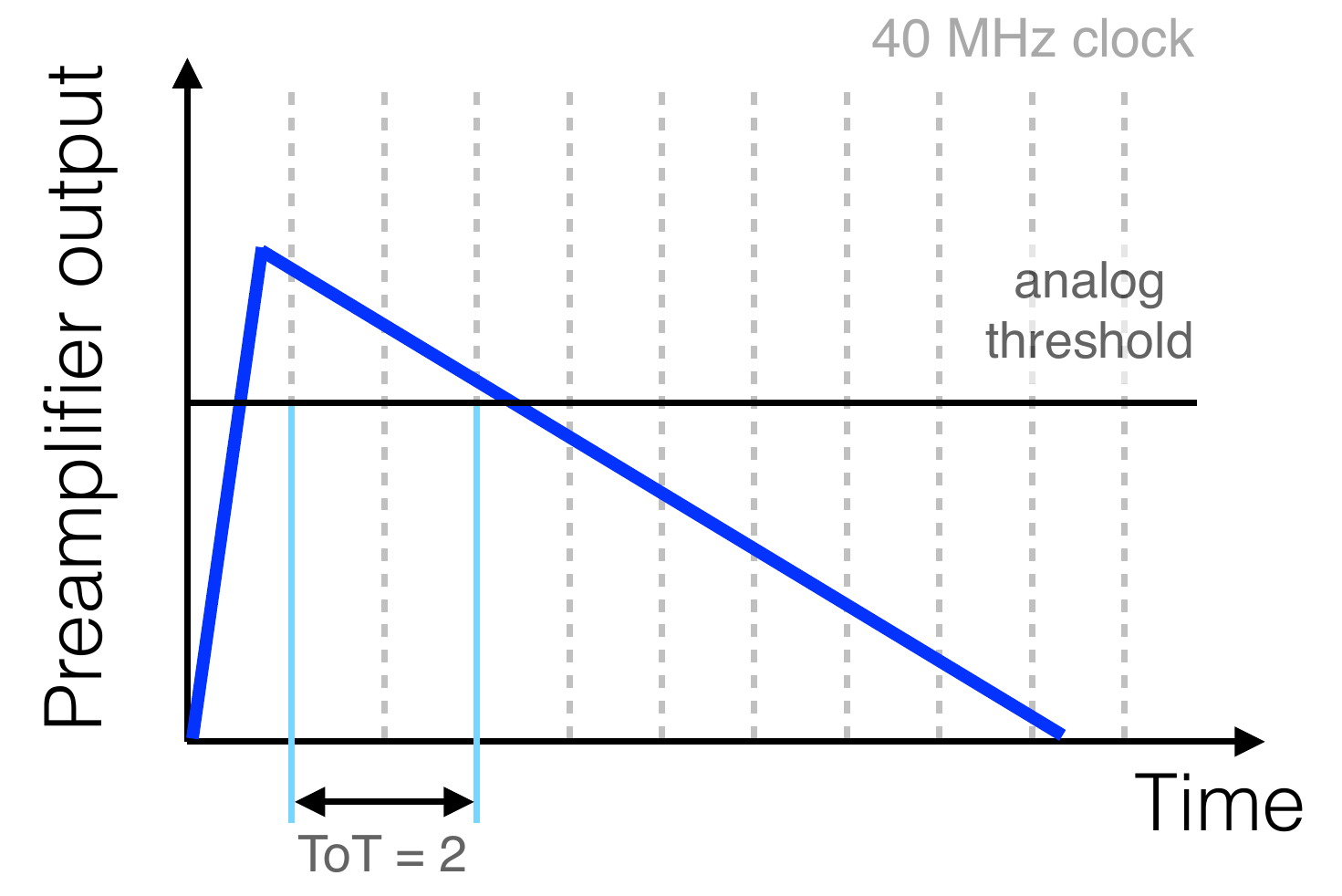}
\caption{A schematic illustration of the charge digitization for a single particle.}
\label{fig:ToTschematic}
\end{figure}

This paper is organized as follows.  An overview is given in Sec.~\ref{sec:tot} and the simulations using these schemes for various studies are documented in Sec.~\ref{sec:sim}.  The subsequent sections analyze the impact of charge digitization on hit efficiency (Sec.~\ref{sec:effic}), charged particle cluster classification (Sec.~\ref{sec:classification}), position resolution (Sec.~\ref{sec:resolution}), and particle identification (Sec.~\ref{sec:classification}).  The paper ends in Sec.~\ref{sec:concl} with conclusions and future outlook.

\section{Overview}
\label{sec:tot}

We assume charge is digitized using the ToT method, as in the RD53A chip, as ToT features 
low power and minimal required circuitry. The pixel amplifier is a charge integrator, which is followed by 
a threshold discriminator. The pixel signal is collected and integrated in approximately 10\,ns and 
a continuous reset of the integrator provides a near linear discharge with adjustable slope of a few 100\,ns per 
Minimum Ionizing Particle (MIP) average charge deposit of 80 electron-hole pairs per micron of path length in silicon ($Q_\text{MIP}$). 
The ToT method measures the time that the discriminator is above threshold using a counter running with a constant frequency clock. The counter starts when the discriminator fires and stops when it returns below threshold. 
The clock is typically that of the accelerator bunch crossings (40 MHz), as chip operation is already synchronized to it.  However, there are various possibilities for increasing the clock speed such as using both the leading and trailing edge of the clock signal to double the frequency.   The ToT value is given 
by the discharge time divided by the clock period. The dynamic range is 2$^m$ counts given a counter of $m$ bits. 
The highest ToT value includes overflow for large deposited charge (discriminator still above threshold when counter 
reaches maximum).  

The hit rate limits the discharge rate.
A non-negligible probability for a second hit to arrive while the first hit is still discharging
gives rise to \textit{in-pixel pileup}, which causes the second hit to be lost and the first hit to 
have a wrong charge measurement. 
This in-pixel pileup must be $\lesssim 1\%$ for accurate track and vertex reconstruction at high luminosity. 
This requirement sets a maximum ToT per $Q_\text{MIP}$ in units of time. The translation of time into bits depends 
on the counting rate. For example, given a 200\,ns $Q_\text{MIP}$ discharge time, a 40\,MHz counting rate translates into 
8 ToT counts per $Q_\text{MIP}$, so a 5-bit counter would have 4 $Q_\text{MIP}$ dynamic range.

The conversion from charge to a final digital value that is transferred off-detector has two levels.  The first level is the per-pixel ADC from the ToT.  The second level is from representing the ToT value on-chip and is limited by the buffer size.  Hits are read out from the detector only if a trigger decision is received.  Trigger latency can be a few microseconds and so the chips must be able to store many hits in a buffer.  The more space allocated to charge information, the fewer hits can be stored on chip.   

So far we have discussed a linear ADC scale. 
However, since the charge deposition probability distribution function is not uniform,
a non-linear scale can use the available bits more efficiently. We therefore explore the performance of non-linear
ADC scales for $n$ stored ToT bits per hit per pixel\footnote{A common memory block may be used for multiple pixels; exploiting the non-uniformity of charge also in space can lead to further efficiency gains.}. 
Regardless of how it is implemented, a non-linear ADC scale can be represented as a mapping from a linear 
scale with $M$ divisions to a storage code that can take on $N$ values\footnote{A non-linear scale can be seen as a specific form of data compression which can be economical to implement and which uses a fixed number of bits for the compressed data.}.
Typically, $M$ and $N$ are powers of 2, because counters and storages area are both binary: 
$M=2^m$ for an $m$-bit binary counter and $N=2^n$ for an $n$-bit storage register. 
Any mapping of interest for this application should map the first (last) ADC element to the first (last) storage element:
$0 \rightarrow 0$  and $(M-1) \rightarrow (N-1)$, it should use all the storage codes at least once,
 and it should be monotonic (element $i+1$ represents a value larger than element $i$). With these
assumptions there are ${M-1} \choose {N-1}$ possible mappings (App.~\ref{sec:app}). We investigate all possible mappings 
in Sec.~\ref{sec:resolution}, but in general we explore two well-motivated non-linear ADC functions called \textit{Kink} and \textit{Exponential},
as described in Sec.~\ref{sec:sim}. 

We expect no scheme can be universally optimal because different uses of charge information have conflicting requirements. 
Three competing metrics for optimizing the use of charge information are hit efficiency (Sec.~\ref{sec:effic}), 
position resolution  (Sec.~\ref{sec:resolution}), 
and particle identification (Sec.~\ref{sec:classification} and~\ref{sec:pid}). 
The left plot of Fig.~\ref{fig:ToTschematic} shows a diagram of a pixel pre-amplifier output with the pileup of two hits,
as explained in Sec.~\ref{sec:intro}. A fast discharge with low charge resolution is useful to avoid in-pixel pileup but is sub-optimal for determining cluster properties. 
The middle plot of Fig.~\ref{fig:ToTschematic} illustrates charge collection in multiple pixels traversed by a MIP.  
As the charge is proportional to the path length in silicon, the location of the entry position is sensitive to small amounts of charge deposited in the first pixel, while measuring high charge values is not important. 
Finally, the right plot of Fig.~\ref{fig:ToTschematic} shows the distribution of energy loss for one and two simultaneous MIPs in the same pixel.  The deposited charge can be used to determine the number and/or type of particles traversing a single pixel cluster. 
In contrast to the low charge sensitivity required by the position resolution, particle multiplicity and type identification benefit from high dynamic range to resolve large values of charge.  
Each of these metrics are quantitatively studied with numerical results based on the pixel simulations described 
Sec.~\ref{sec:sim}.

\begin{figure}[h!]
\centering
\includegraphics[width=0.95\textwidth]{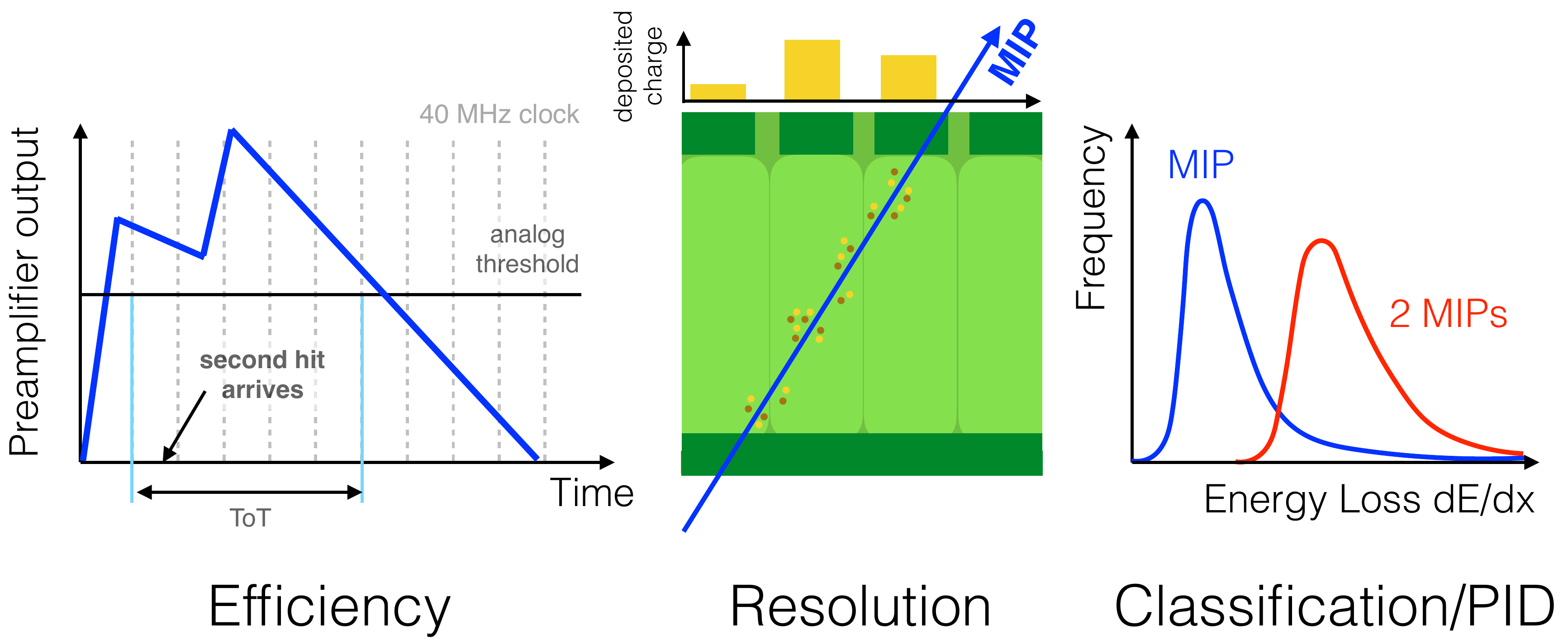}
\caption{Three diagrams representing the three (competing) metrics for optimizing the use of charge information.  Left: the preamplifier output as a function of time when a second hit arrives before the signal from the first hit has gone below threshold.  Middle: the energy deposited by a MIP as it traverses several pixels.  Right: the probability distribution for the energy loss for one and two MIPs.}
\label{fig:ToTschematic}
\end{figure}

\section{Simulation}
\label{sec:sim}

The detector description and run conditions are simulated using Allpix~\cite{benoit:20xx}, which is built on the Geant4 package~\cite{Agostinelli:2002hh}.   The ionization energy provided by Geant4 with the \textsc{emstandard\_opt0} model is converted into electron-hole pairs which are transported through the silicon including drift and diffusion.  Energy depositions at a particular location are divided into pieces, each with the same starting position and an energy of about 3.6 eV, the $e$-$h$ ionization energy in silicon.  Each piece is individually drifted to the electrode, accounting for diffusion by smearing each of the directions perpendicular to the depth according to a Gaussian distribution with standard deviation given by $r_\text{diff}$:

\begin{align}
\label{eq:rdiff}
r_\text{diff}=\sqrt{2Dt_\text{electrode}},
\end{align}

\noindent where $D$ is the diffusion constant and $t_\text{electrode}$ is the time required to reach the electrode.  The diffusion constant is calculated by the Einstein relation:

\begin{align}
\label{eq:diffconstant}
D=\frac{k_BTr\mu}{e},
\end{align}

\noindent where $k_B$ is the Boltzmann constant, $T$ is the temperature, $\mu$ is the mobility, $r$ is the Hall scattering factor, and $e$ is the electron charge.  The time required to reach the electrode is calculated using the electron mobility\footnote{In these $n$-in-$p$ sensors, electrons are collected.  Holes also contribute to the transient signal, but are not relevant for the present analysis.}:

\begin{align}
\label{eq:difftime}
t_\text{electrode} = \int_{z_0}^{z_\text{electrode}}\frac{\text{d}z}{r\mu(E(z))E(z)}.
\end{align}

\noindent The mobility is parameterized using the common Canali-modified Caughey and Thomas velocity saturation model~\cite{Caughey1967,Canali1975}:

\begin{align}
\mu(E)=\frac{\mu_0}{\left(1+\left(\frac{\mu_0E}{v_\text{sat}}\right)^\beta\right)^{1/\beta}},
\end{align}

\noindent where the low field mobility $\mu_0(T)=1533.7$ cm$^2/(\text{V}\cdot\text{s})\times(T/\text{300 K})^{-2.42}$, the saturation velocity is given by $v_\text{sat}=1.07\times 10^7$ cm/s$\times (T/\text{300 K})^{-0.87}$, and $\beta=1.109\times(T/\text{300 K})^{0.66}$.   The temperature is set to $273$ K.  In addition to diffusion, charges are deflected in a 2 T magnetic field that is perpendicular to the sensor depth.  The angle of deflection is the Lorentz angle, given by $\tan\theta =r \mu B$. 

To model ATLAS and CMS-like sensors for the HL-LHC, pixels have dimensions 50 x 50 x 150 $\mu$m$^3$ and are placed a distance of 3 cm from the interaction point.  The challenges facing pixel sensors are most severe for the innermost layer.  The number of ToT bits is scanned between $3$ and $6$ using three different schemes for the charge to ToT conversion\footnote{The conversion is modeled as uniform across the pixel matrix.  In practice, there is a small spread in the tuning that can be partially corrected for with per-pixel calibration curves in offline analysis.  Any residual degradation is beyond the scope of this study.}.  In the first scheme, ToT is proportional to the amount of charge over threshold.  The proportionality constant is usually specified by the ToT ($\text{ToT}_\text{MIP}$) of the charge expected by a MIP at perpendicular incidence, $Q_\text{MIP}$.  Unless otherwise specified, $\text{ToT}_\text{MIP}$ will be half of the available range, $\text{ToT}_\text{half} = 2^{n-2}$. Two additional schemes are constructed to enhance the sensitivity to large and extreme amounts of deposited charge, while maintaining some precision at low charge.  Both schemes are the same as the linear scheme up to $\text{ToT}_\text{half}$.  Section~\ref{sec:resolution} will show that there is little resolution information for charges beyond the MIP peak which is why this is the point where the non-linearity begins.  The first additional scheme is also linear beyond the MIP peak, but with a shallower slope than for the first $\text{ToT}_\text{half}$ values.  This increases the sensitivity to high charge.  A third scheme extends this idea further by using a ToT that is logarithmic in charge to extend precision to charges from many MIPs.


Table~\ref{tab:chargeToToTschemes} summarizes the three charge to ToT schemes studied in this paper.   The important feature of the kink and exponential schemes is that they probe higher charges than the usual linear scheme.  The kink scheme slope doubles past $Q_\text{MIP}$ and the charge range corresponding to one ToT bit after the $Q_\text{MIP}$ for the exponential scheme doubles for each bit.  Current readout limitations limit also place constraints on the maximum charge that can be measured.  For example, for pixels with 50 fF capacitance, a charge of 50 fC will cause a 1V potential increase, reaching the maximum value typical readout chips can handle.   For 80 electrons/$\mu$m over 150 $\mu$m, $Q_\text{MIP}\approx 2$ fC; therefore, in the exponential scheme, beyond $25\times Q_\text{MIP}$, the ToT is fixed at $25\times Q_\text{MIP}$.  

\begin{figure}[h!]
\centering
\includegraphics[width=0.55\textwidth]{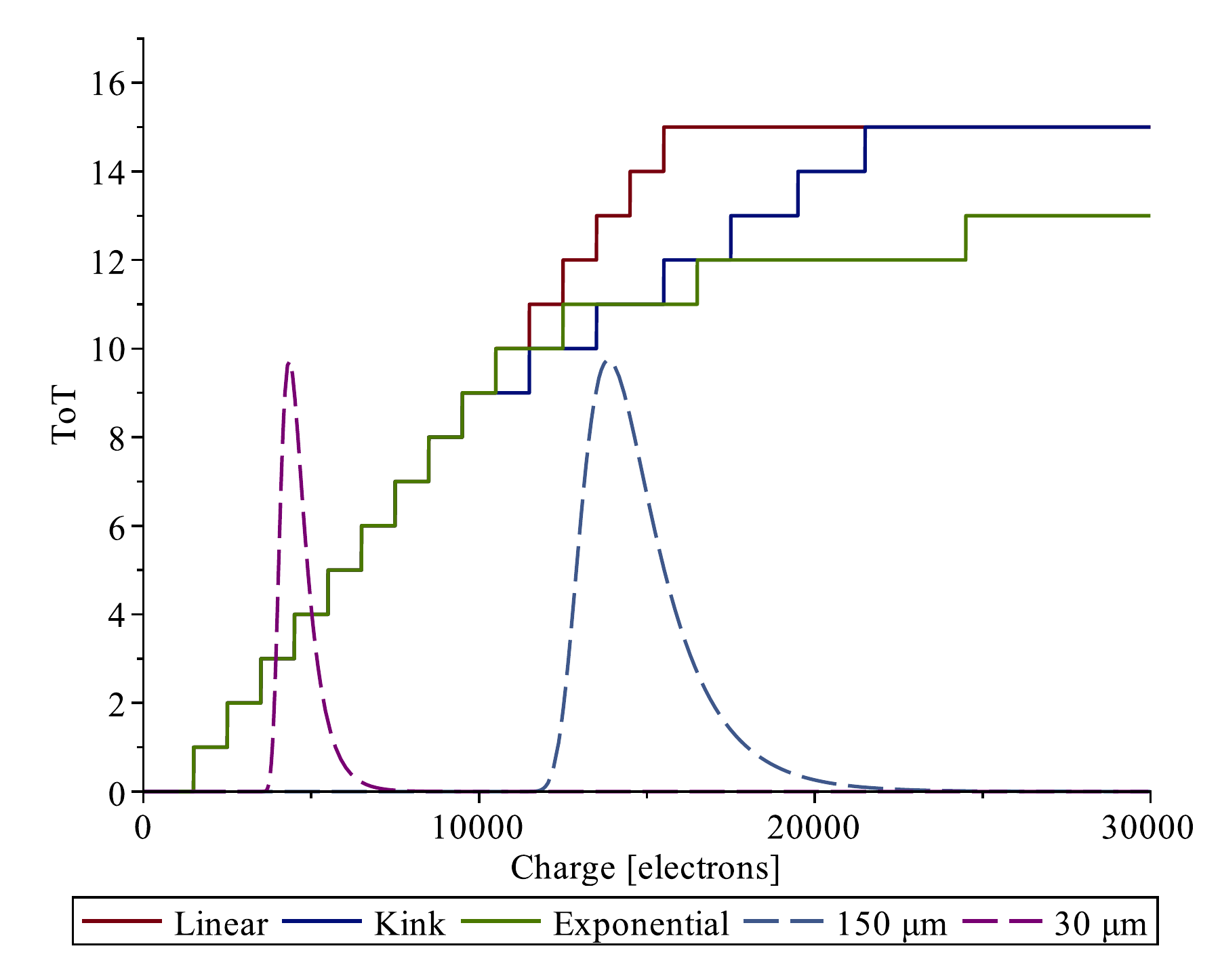}
\caption{An illustration of the three charge to ToT schemes studied in this paper.  In the linear scheme, ToT is proportional to charge; for the kink scheme, ToT is proportional to charge, but the proportionality constant changes after $\text{ToT}_\text{half}$; for the exponential scheme, ToT is proportional to charge until $\text{ToT}_\text{half}$, after which the charge range covered by a bit of ToT doubles with increasing ToT values.  For illustration, the Landau component of the charge fluctuations for a MIP traversing 30 or 150 $\mu $m of silicon are shown.}
\label{fig:ChargeToToTSchemes.pdf}
\end{figure}

Figure~\ref{fig:ChargeToToTSchemes.pdf} illustrates the charge to ToT conversion.  By construction, the three methods are identical below $Q_\text{MIP}$ and cover different charge ranges beyond the peak.  For comparison, the expected charge distribution is shown for a MIP passing through the full sensor (150 $\mu$m) as well as only a fraction (20\%) of the sensor.  Ideally the charge calibration would be $\eta$-dependent, since for $|\eta|\gtrsim 1$, particles originating from the geometric center of the detector would pass through a pitch's worth of silicon instead of the depth.  Low charges are also natural at all $\eta$ for the edge pixels in a cluster. 

\begin{table}[h!]
\centering
\noindent\adjustbox{max width=\textwidth}{
\begin{tabular}{| c | c | c |}
  \hline
  Method & Functional Form & Maximum Charge*  \\
  \hline\rule{0pt}{5ex}
  Linear & $\text{ToT}_\text{L}(Q)=\text{min}\{\floor*{\alpha\times \text{max}\{(Q-T),0\}},2^{n}-2\}+1- \mathbbm{1}(Q<T) $ & $\frac{1}{\alpha}(2^n-2)+T$  \\\rule{0pt}{5ex}
   Kink &$\text{ToT}_\text{K}(Q)=\left\{\begin{matrix}\text{ToT}_\text{L}(Q) & Q < Q_\text{MIP}+T \cr \text{min}\{\floor*{\frac{1}{2}\alpha\times (Q-T-Q_\text{MIP})}+2^{n-1}+1,2^{n}-1\} & \text{else}\end{matrix}\right.$  & $\frac{1}{\alpha}(2^{n+1}-2^n+2^{n-1}-4)+T$\\\rule{0pt}{5ex}
    Exponential & $\text{ToT}_\text{E}(Q)=\left\{\begin{matrix}\text{ToT}_\text{L}(Q) & Q < Q_\text{MIP}+T \cr\text{min}\{\floor*{\log_2(1+\alpha\times (Q-T-Q_\text{MIP}))}+2^{n-1}+1,2^{n}-1\} & \text{else}\end{matrix}\right.$ &$\frac{1}{\alpha}(2^{2^n-2^{n-1}-2}+2^{n-1}-1)+T$ \\
      \hline
\end{tabular}}
\caption{The three schemes for converting charge to ToT studied in this paper.  The parameter $\alpha = \text{ToT}_\text{MIP}/Q_\text{MIP}$.  The last column is the maximum charge that is not in the overflow bin.  The symbols $n$ and $T$ stand for the number of bits and the analog charge threshold, respectively. *For the exponential scheme, the ToT is also capped at $25\times \text{ToT}_\text{MIP}$; see the text for details.}
\label{tab:chargeToToTschemes}
\end{table}

\begin{figure}[h!]
\centering
\includegraphics[width=0.55\textwidth]{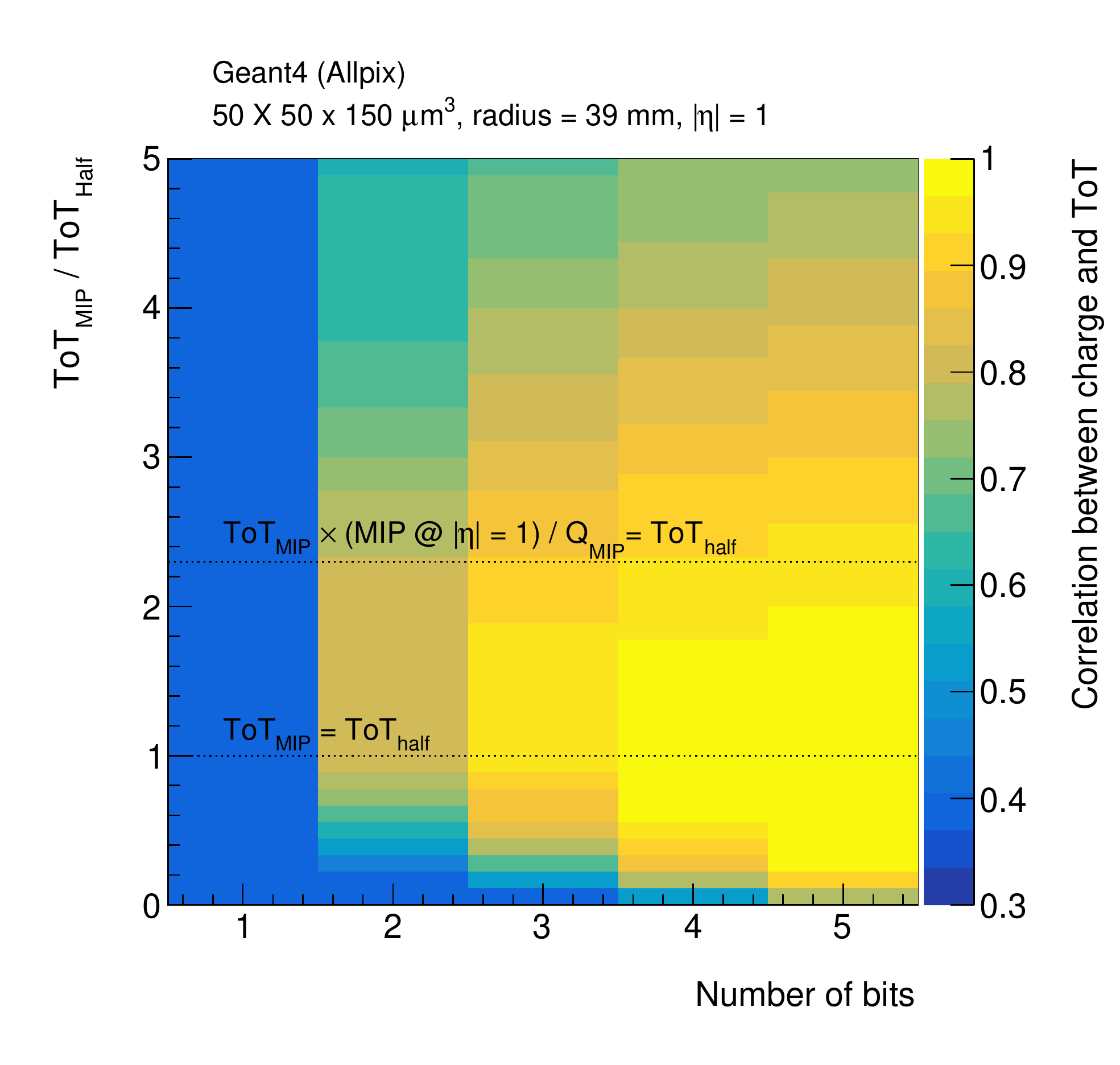}
\caption{The linear correlation between the charge and the ToT as a function of the number of bits and $\text{ToT}_\text{MIP}$, relative to $\text{ToT}_\text{half}$.  To guide the eye, typical tuning values are indicated that correspond to $\text{ToT}_\text{MIP}=\text{ToT}_\text{half}$ and to where the actual MIP peak for $|\eta|=1$ correspond to half of the available ToT range.  Recall that the symbol $Q_\text{MIP}$ corresponds to 80 electrons/$\mu$m at perpendicular incidence.}
\label{fig:correlation}
\end{figure}

Our goal is to quantify the performance of three charge to ToT schemes for a scan in the number of ToT bits.  Before focusing on the tasks described in the previous section, Fig.~\ref{fig:correlation} shows the impact of charge discretization.   As expected, the correlation\footnote{The linear correlation is defined as the covariance divided by the product of standard deviations.} between the charge and the ToT increases with more bits.  When the MIP ToT is too high or too low, the large over/underflow fractions reduce the correlation.  The correlation is already nearly one for 4 bits and a MIP peak at half of the available range.  The linear correlation is a crude metric for quantifying the information content of the ToT, which is investigated in more detail in the subsequent sections.

\clearpage

\section{Efficiency}
\label{sec:effic}

\begin{figure}[h!]
\centering
\includegraphics[width=0.8\textwidth]{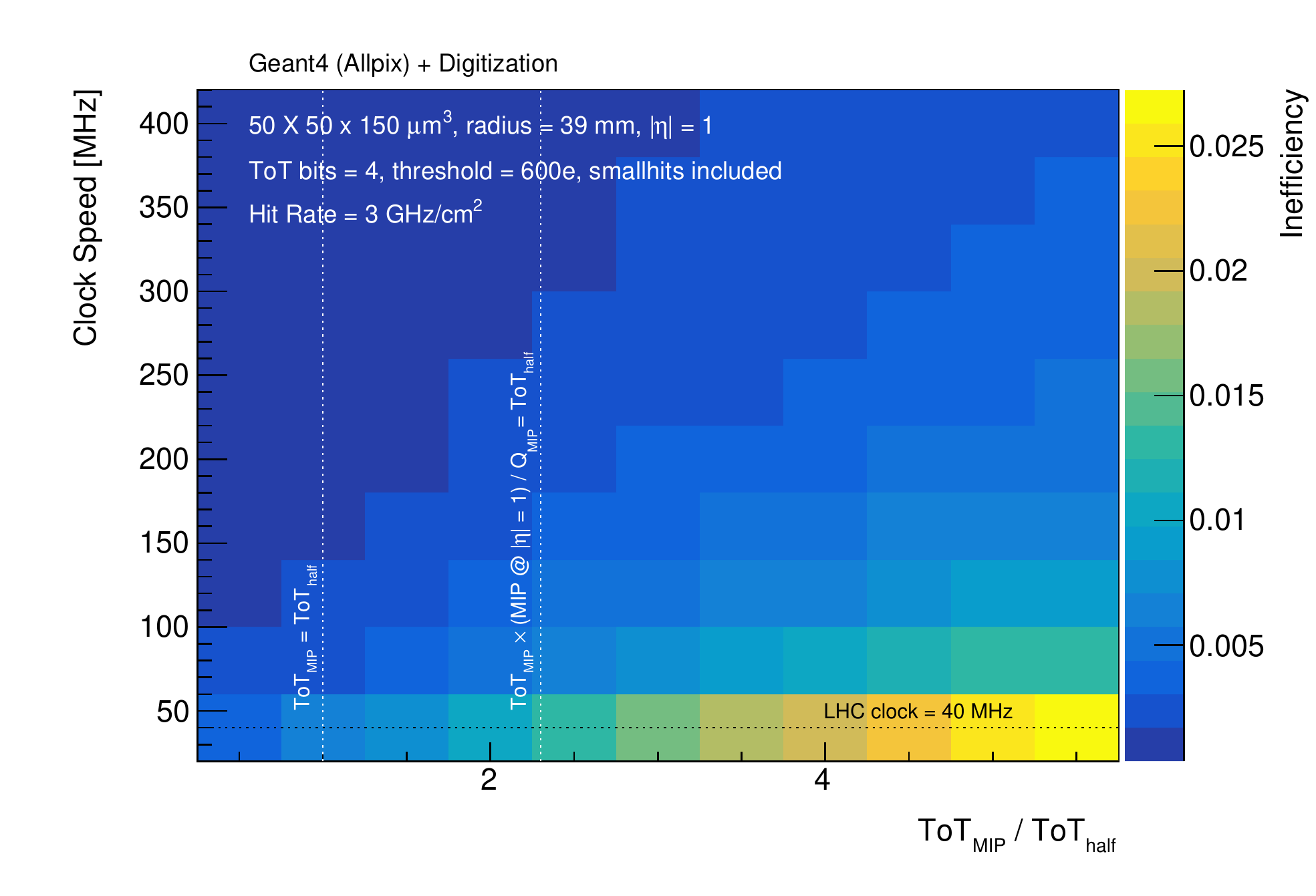}
\caption{The probability for in-pixel pileup in the linear charge to ToT scheme as a function of $\text{ToT}_\text{MIP}$ and the clock speed.  The nominal LHC clock is 40 MHz. To guide the eye, typical tuning values are indicated that correspond to $\text{ToT}_\text{MIP}=\text{ToT}_\text{half}$ and to where the actual MIP peak for $|\eta|=1$ correspond to half of the available ToT range.  Recall that the symbol $Q_\text{MIP}$ corresponds to 80 electrons/$\mu$m at perpendicular incidence.}
\label{fig:efficiency:fasttiming}
\end{figure}

One can achieve a high precision across the entire charge spectrum by forcing a very slow discharge and counting for a long time.  However, this will not work in practice because the probability for a second hit to arrive before the preamplifier output goes below threshold from the first hit will be large at the HL-LHC.  In particular, at the HL-LHC the hit density will be approximately $r\approx3$ GHz/cm$^{2}$ (RD53 spec), which is 0.2\% occupancy per pixel per BC for a $50\times 50$ $\mu$m$^2$ pixel.  One possibility is to count faster than 40 MHz, as in the case of the synchronous RD53A front-end~\cite{7581969}.   Figure~\ref{fig:efficiency:fasttiming} shows the probability for in-pixel pileup in the linear charge to ToT scheme as a function of $\text{ToT}_\text{MIP}$ and the clock speed.  Adding a faster clock requires additional logic and requires careful study to ensure reliability.  Figure~\ref{fig:efficiency:fasttiming} shows that if these challenges are addressed then a $\lesssim 200$ MHz clock will reduce in-pixel pileup (well) below $1\%$ for tuning parameters in the relevant regime ($\text{ToT}_\text{MIP}\lesssim 3\times \text{ToT}_\text{half}$).

For a fixed clock speed, one can always reduce the in-pixel pileup by tuning the discharge to happen very quickly.  However, this is an inefficient use of the ToT counter and reduces the charge precision.  Therefore, it is likely that the charge distribution from a MIP will have a non-negligible contribution in the ToT overflow bit.  Any information in charge values above the overflow value are lost.  One way to mitigate this inefficiency may be to quickly reset the integrator when the ToT counter reaches maximum (overflow).  For a given hit rate $r$, one can compute the inefficiency due in-pixel pileup resulting from overflow using the charge distribution $\Pr(\text{ToT})$: 

\begin{align}
\label{eq:fastdischarge}
1-\epsilon &= \sum_{\text{ToT}=2^{n}-1}^\infty \Pr(\text{ToT})\times\Pr(\text{2$^\text{nd}$ hit at or before ToT BCs})\\\nonumber
&=\sum_{\text{ToT}=2^{n}-1}^\infty \Pr(\text{ToT})\times(1-e^{-\text{ToT}\times r}).
\end{align}

\begin{figure}[h!]
\centering
\includegraphics[width=0.8\textwidth]{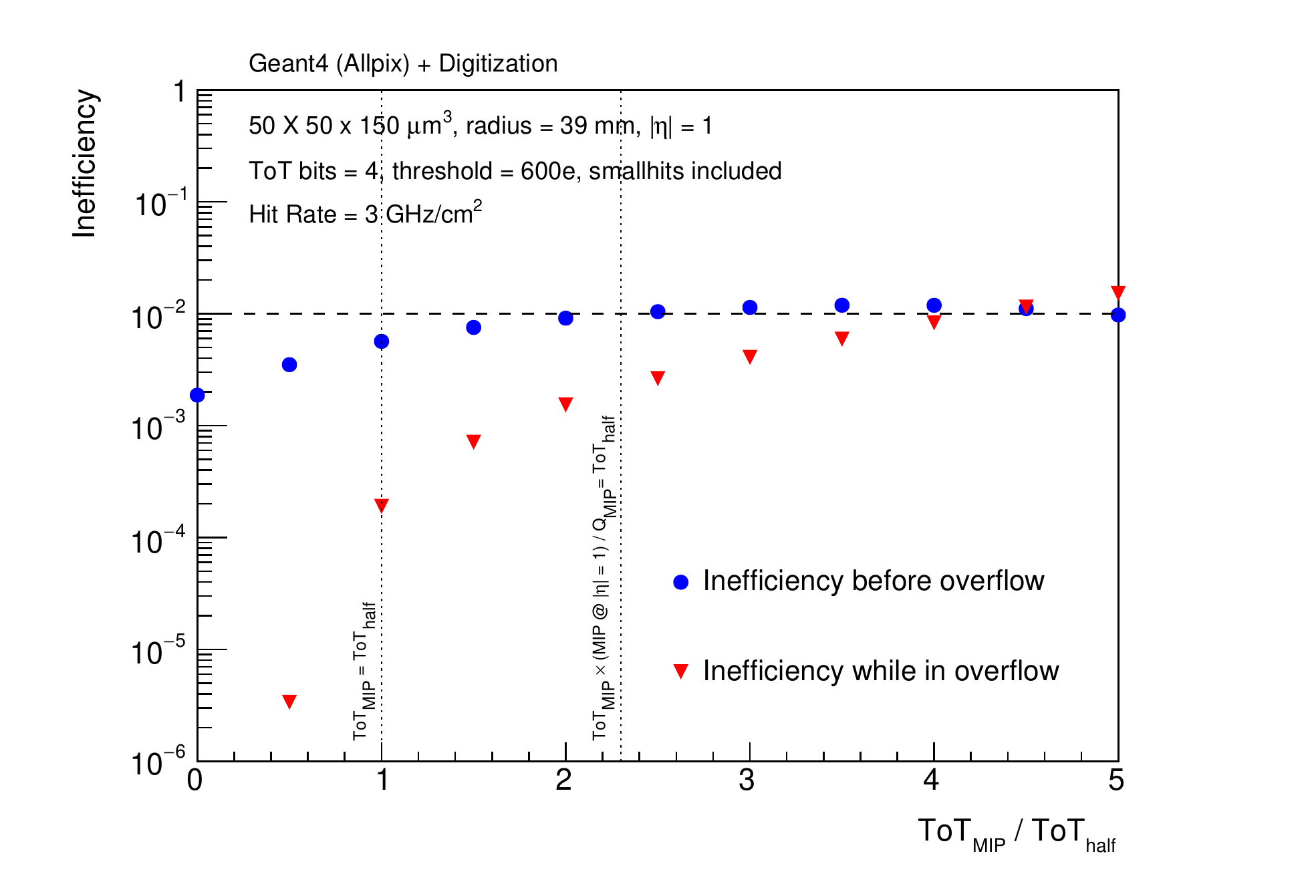}
\caption{For a 3 GHz/cm$^2$ hit rate, the blue circles show the probability for a second hit to arrive during the first $\text{ToT}_\text{half}$ values and the red shows the probability that a second hit arrives while the counter is in overflow, but still above threshold.  To guide the eye, typical tuning values are indicated that correspond to $\text{ToT}_\text{MIP}=\text{ToT}_\text{half}$ and to where the actual MIP peak for $|\eta|=1$ correspond to half of the available ToT range.  Recall that the symbol $Q_\text{MIP}$ corresponds to 80 electrons/$\mu$m at perpendicular incidence.}
\label{fig:efficiency:fastdischarge}
\end{figure}

\noindent The inefficiency described by Eq.~\ref{eq:fastdischarge} can be completely removed with a fast reset scheme.  The triangles in Fig.~\ref{fig:efficiency:fastdischarge} shows the overflow inefficiency as a function of the tuning in the linear charge to ToT scheme.  For comparison, Fig.~\ref{fig:efficiency:fastdischarge} also shows the contribution to in-pixel pileup from additional hits arriving before the ToT counter reaches overflow.  In the regime near $\text{ToT}_\text{MIP}\sim \text{ToT}_\text{half}$, the overflow inefficiency is small compared to the non-overflow inefficiency and therefore there is likely not a significant gain from adding a fast reset component to the chip.

\begin{figure}[h!]
\centering
\includegraphics[width=0.33\textwidth]{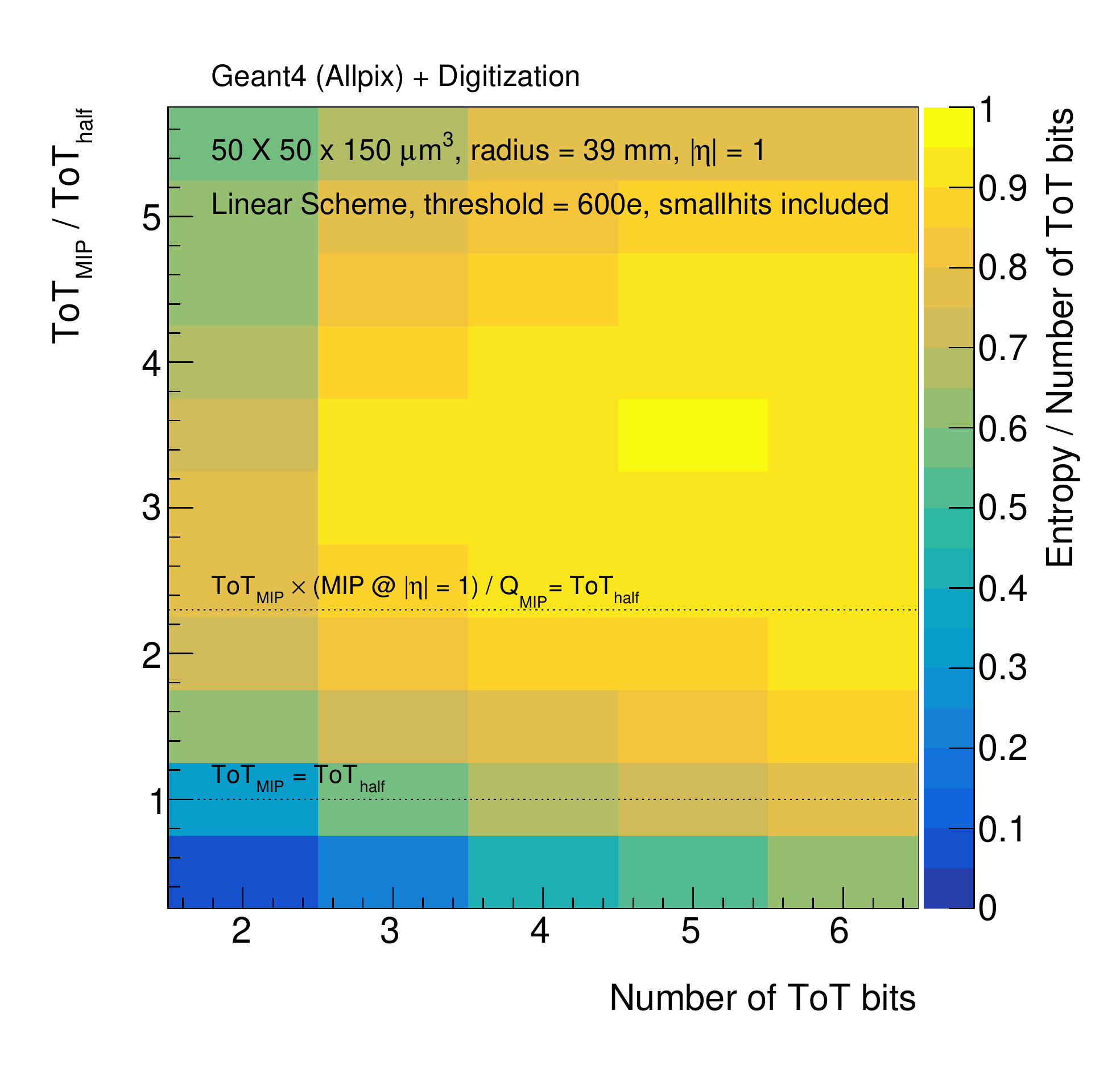}\includegraphics[width=0.33\textwidth]{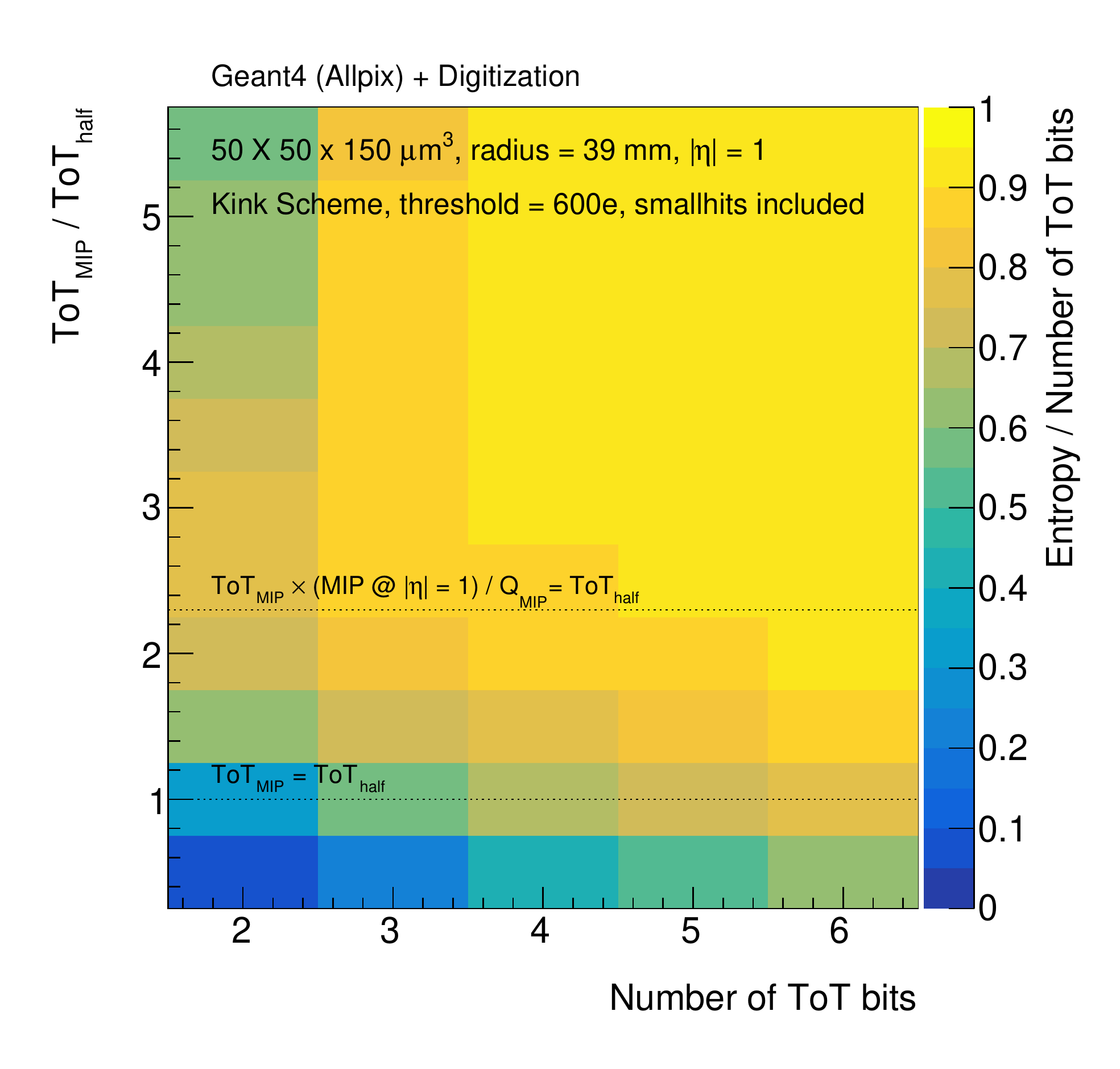}\includegraphics[width=0.33\textwidth]{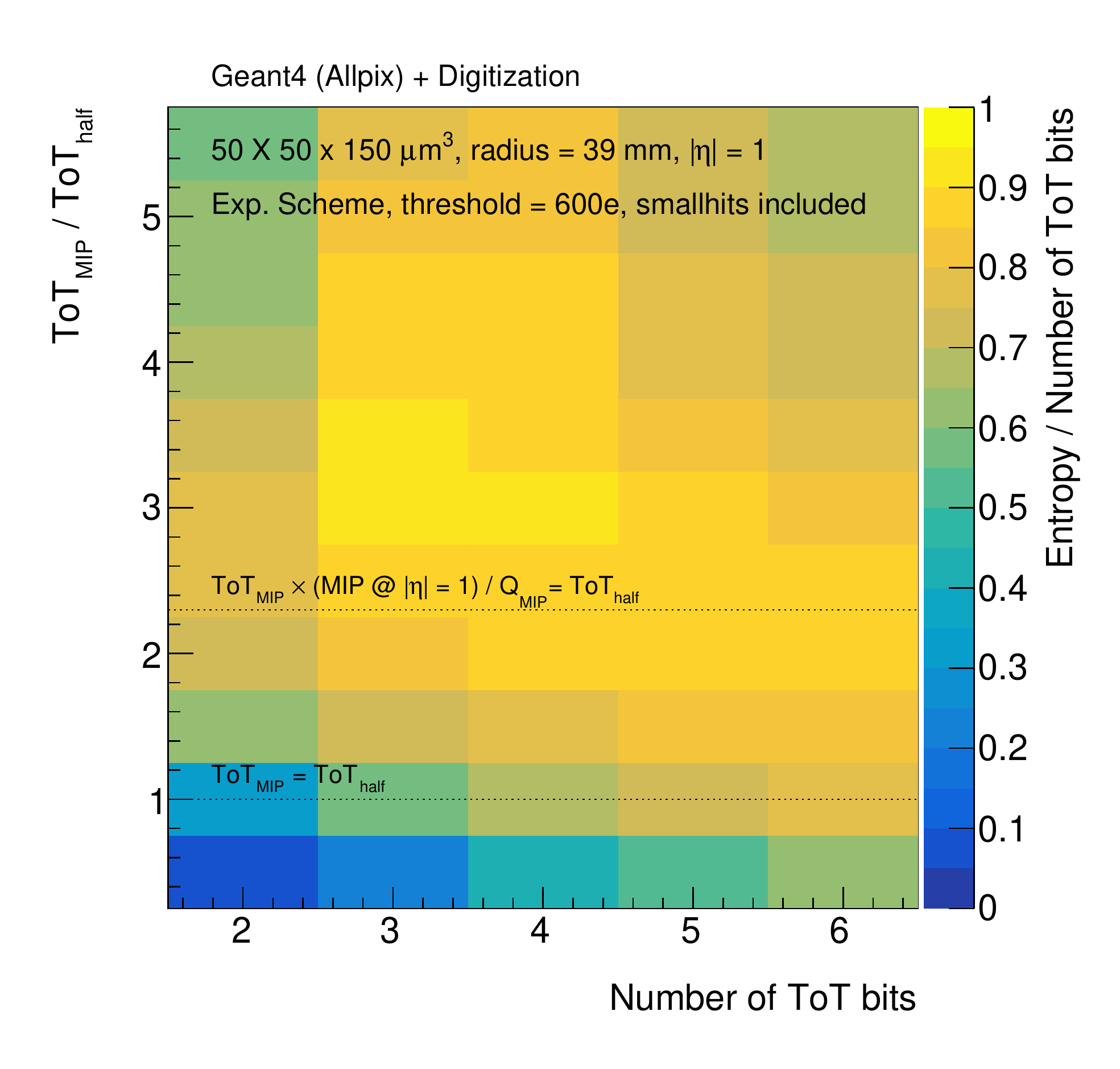}
\caption{The entropy divided by the number of ToT bits as a function of the number of ToT bits and the tuning for the linear (left), kinked (middle), and exponential (right) charge to ToT schemes. To guide the eye, typical tuning values are indicated that correspond to $\text{ToT}_\text{MIP}=\text{ToT}_\text{half}$ and to where the actual MIP peak for $|\eta|=1$ correspond to half of the available ToT range.  Recall that the symbol $Q_\text{MIP}$ corresponds to 80 electrons/$\mu$m at perpendicular incidence.}
\label{fig:efficiency:entropy}
\end{figure}

As already mentioned, compression of ToT could allow to store more hits without sacrificing precision.  One way to reduce overflow errors in the buffer is to compress the ToT.  Since the ToT distribution is not uniform, the average number of bits required to represent an $n$ bit ToT counter is less than $n$.  This is especially true for the kink and exponential schemes in which the high charge values rarely occur.   The information content in units of bits is calculated using \textit{entropy}, which is given by $s_n=-\sum_{i=0}^{2^{n}-1}\Pr(i)\log_2(\Pr(i))$, where $\Pr(i)$ is the probability to measure a ToT value of $i$.  When $s_n<n$, there is an opportunity to gain by compression.  Figure~\ref{fig:efficiency:entropy} shows $s_n/n$ as a function of $\text{ToT}_\text{MIP}$ and $n$.  As expected, when $\text{ToT}_\text{MIP}$ is much less than or much greater than $\text{ToT}_\text{half}$, the $n$ bits are not used very effectively; when $\text{ToT}_\text{MIP}\sim \text{ToT}_\text{half}$, the bits are used effectively (distribution closer to uniform).
 

\clearpage

\section{Classification}
\label{sec:classification}

Charge information is key to identifying the number of particles traversing a single pixel cluster.  The dense core of high $p_\text{T}$ jets, hadronically decaying multi-prong $\tau$ leptons, and conversion photons are all cases where pixel clusters from multiple particles can merge into a single cluster.  On average, the charge deposited by multiple MIPs will be higher than for single MIPs.  This section explores the ToT precision needed to identify the number of particles traversing a pixel cluster.  Determining the position of these particles is further explored in Sec.~\ref{sec:resolution} and Sec.~\ref{sec:pid} extends the multiplicity classification to particle type identification.  To begin, Sec.~\ref{sec:class:onepix} examines multiple particles traversing a single pixel.  By ignoring cluster information, it is possible to quantify what can be accomplished with only ToT for single pixel classification.  The charge deposited in a single pixel does not follow a Landau-like distribution due to charge sharing from diffusion.  In order to remove this effect, the ToT for nearby pixels in the $\phi$ direction is summed.  

\subsection{One pixel at perpendicular incidence}
\label{sec:class:onepix}

\begin{figure}[h!]
\centering
\includegraphics[width=0.3\textwidth]{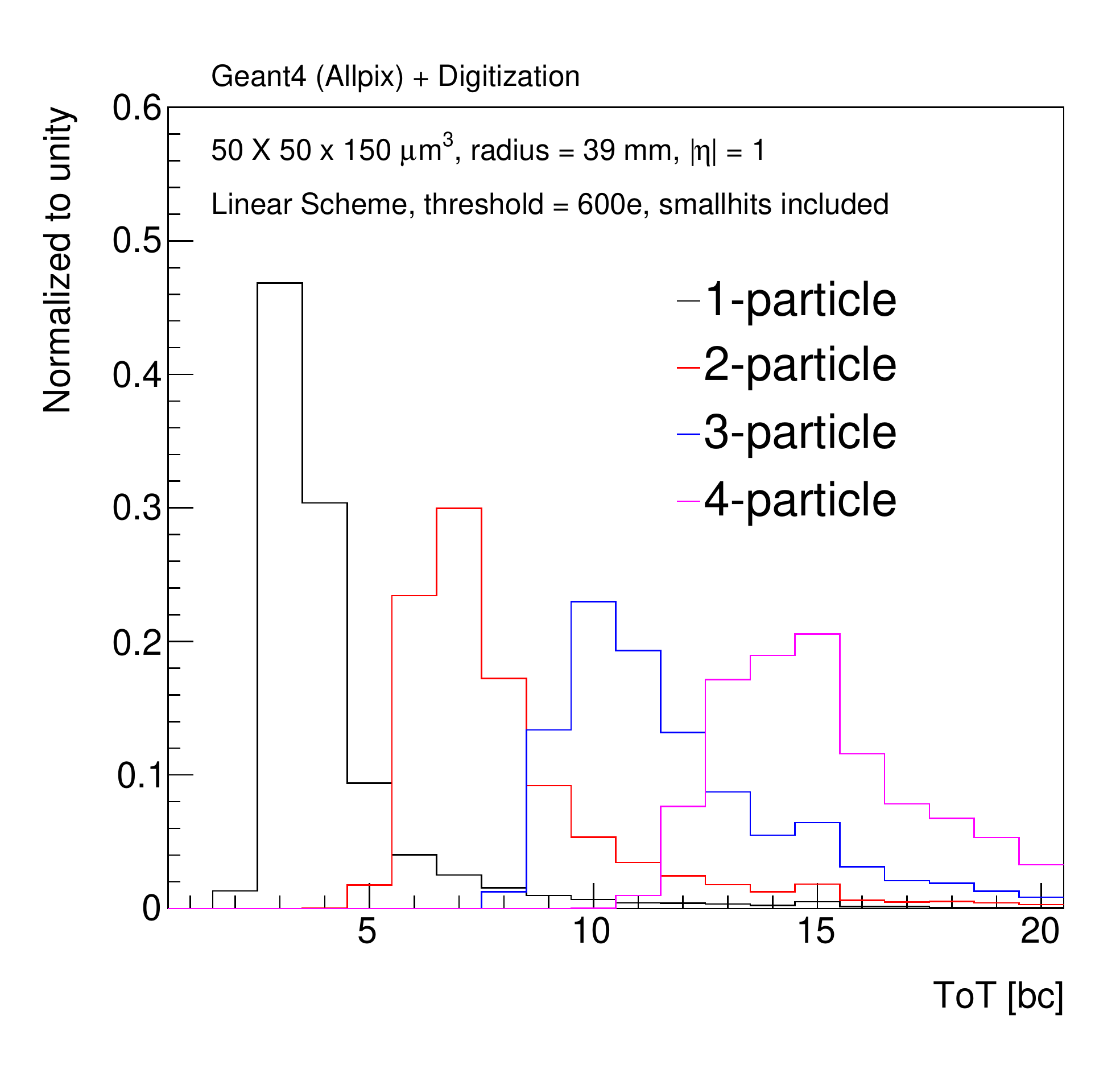}\includegraphics[width=0.3\textwidth]{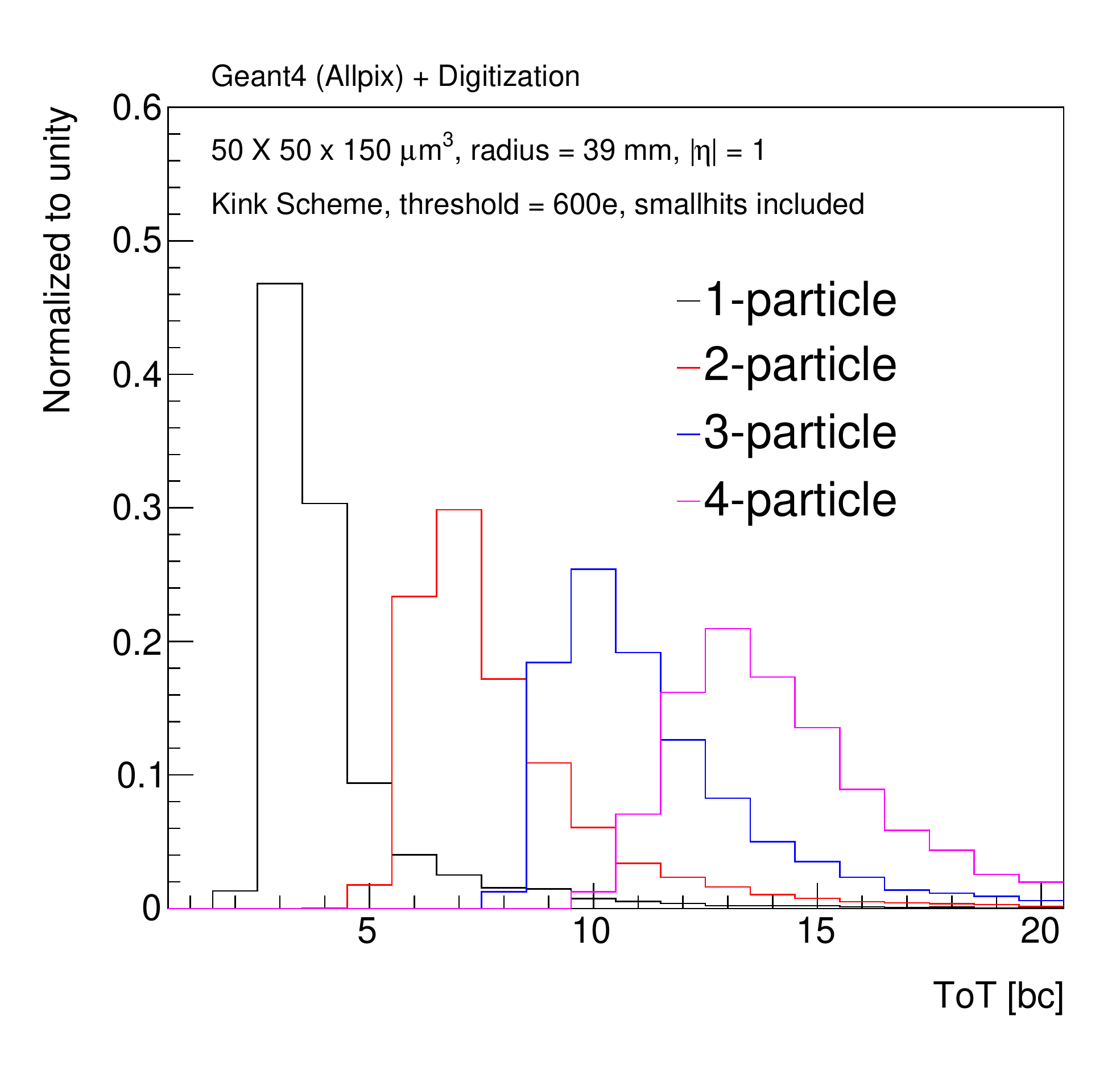}\includegraphics[width=0.3\textwidth]{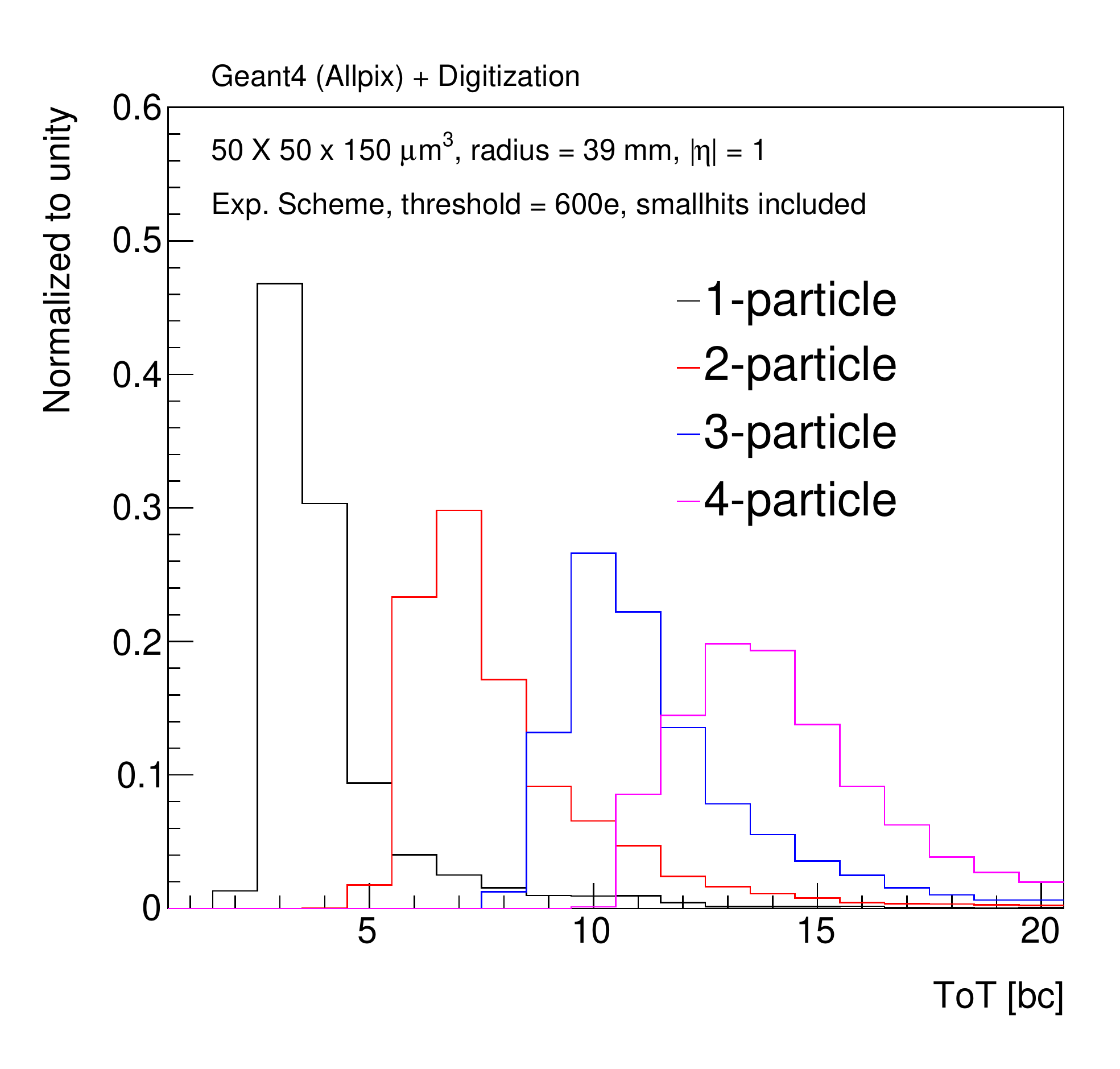}
\caption{The ToT distribution for up to four MIPs traversing a pixel.  The first and last pixel along the $\eta$ direction are excluded and the ToT is summed across pixels in the $\phi$ direction. The left plot uses the linear charge to ToT scheme, the middle uses the kink and the right uses the exponential scheme.}
\label{fig:TwoParticleID:simple1}
\end{figure}

The ultimate particle multiplicity identification uses the full distribution of charge across the cluster as well as cluster shape information.  Before considering the general case in Sec.~\ref{sec:class:general}, this section focuses on using only pixel ToT to identify on a pixel-by-pixel basis how many particles traversed the cluster.  In one dimension, it is possible to do a broad scan of parameters that is computational expensive with the neural network approach described in Sec.~\ref{sec:class:general}.  Figure~\ref{fig:TwoParticleID:simple1} shows the ToT distribution for up to four particles traversing a single pixel.  The first and last pixels in the $\eta$ direction from the cluster are excluded because the charge distribution in those pixels is skewed low due to the reduced path length.  The kink and exponential charge to ToT conversion schemes are much less sensitive than the linear scheme to the overflow bin.

\begin{figure}[h!]
\centering
\includegraphics[width=0.5\textwidth]{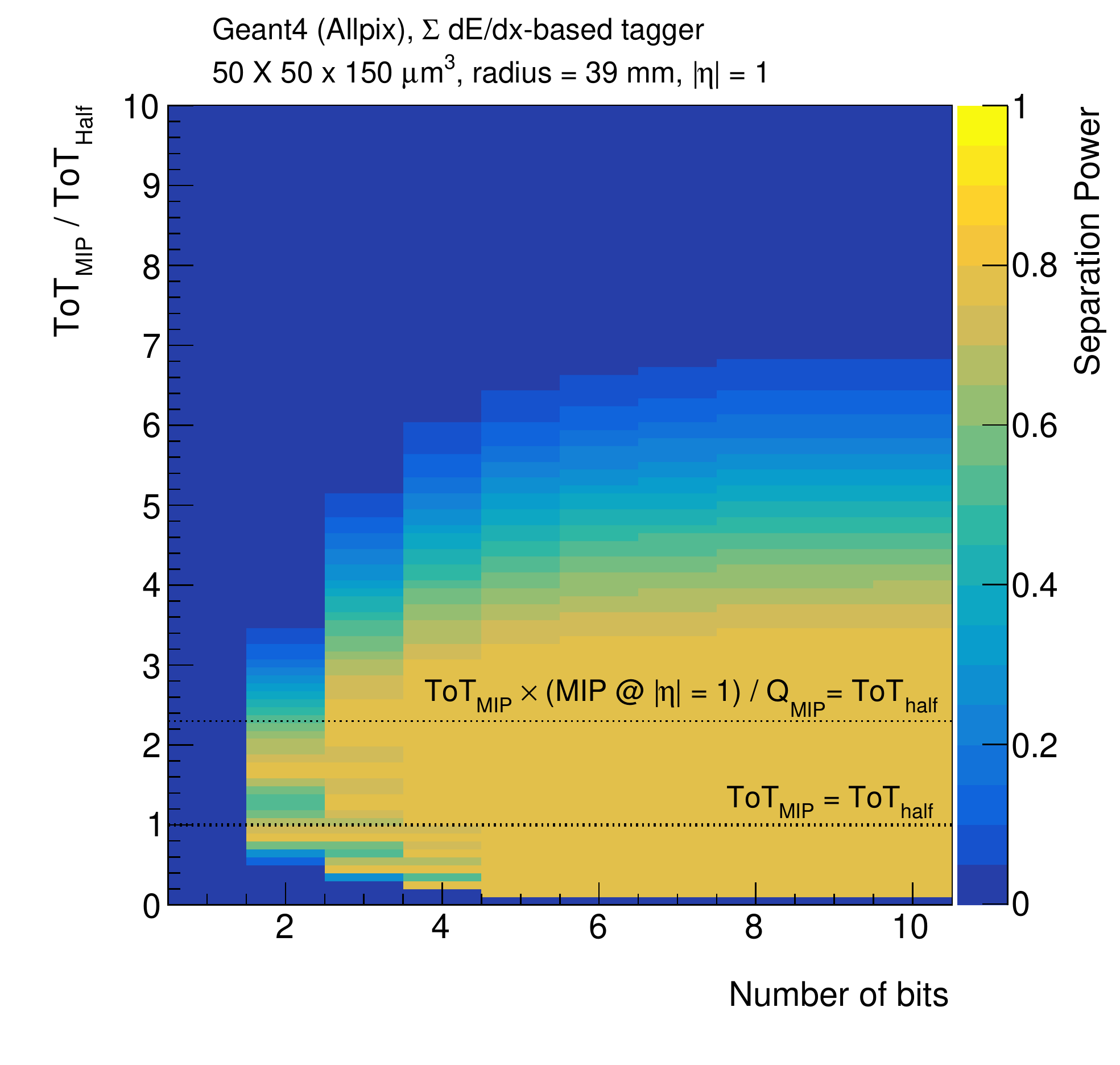}\includegraphics[width=0.5\textwidth]{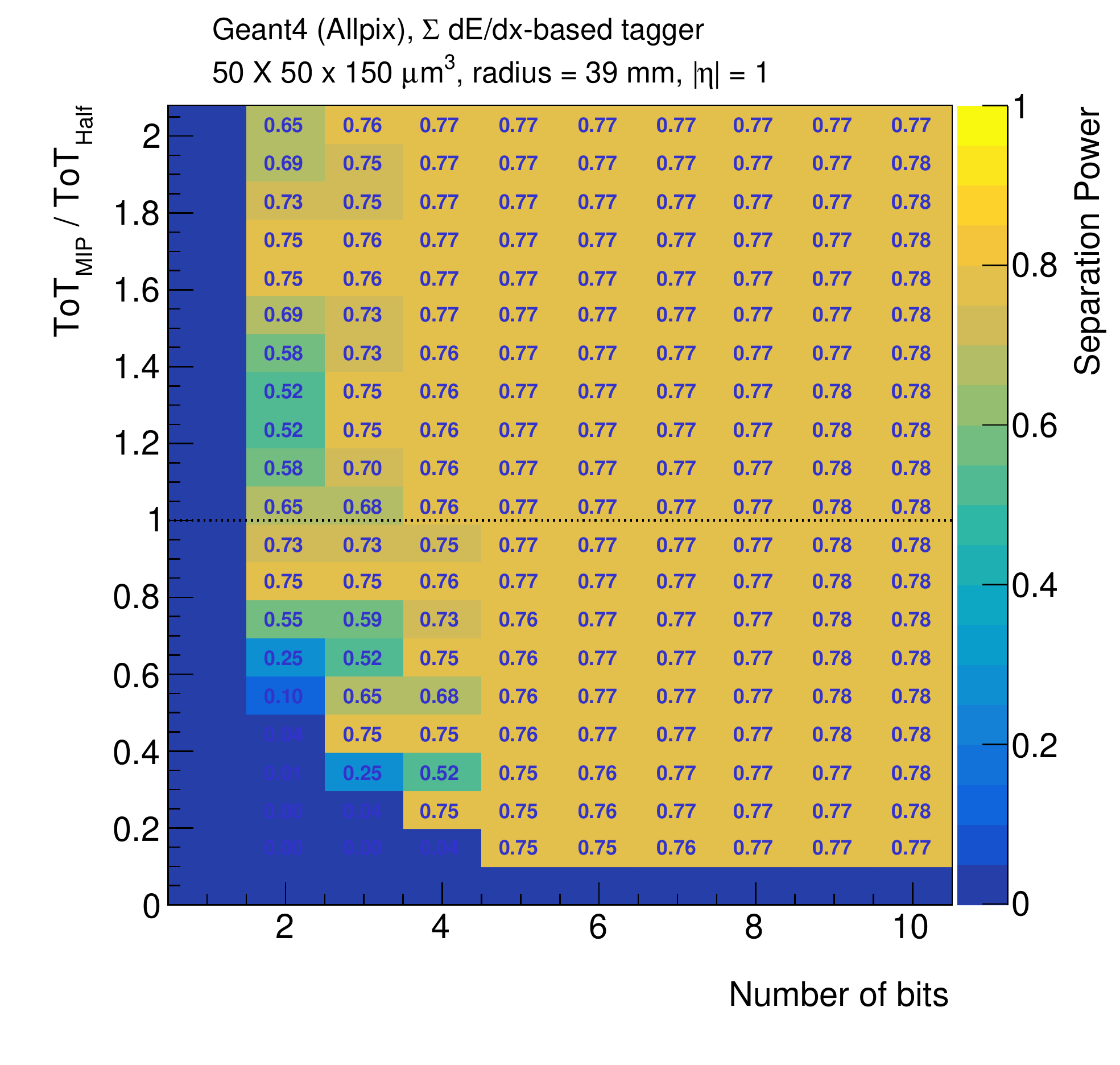}
\caption{The separation power (see Eq.~\ref{eq:seppower}) as a function of the number of ToT bits and the tuning in the linear charge to ToT conversion scheme.  To guide the eye, typical tuning values are indicated that correspond to $\text{ToT}_\text{MIP}=\text{ToT}_\text{half}$ and to where the actual MIP peak for $|\eta|=1$ correspond to half of the available ToT range.  Recall that the symbol $Q_\text{MIP}$ corresponds to 80 electrons/$\mu$m at perpendicular incidence.  The right plot is a zoomed-in version of the left plot.}
\label{fig:TwoParticleID:simple3}
\end{figure}

Due to its importance and for simplicity, the rest of this section focuses on one versus two particle classification.  The more separated the one and two particle distributions are, the easier it is to perform multiplicity classification.  One way to quantify the difference in distributions is with the separation power:

\begin{align}
\label{eq:seppower}
\frac{1}{2}\sum_{i=1}^n \frac{(h_{1,i}-h_{2,i})^2}{h_{1,i}+h_{2,i}},
\end{align}


\noindent where $h_{j,i}$ is the $i^\text{th}$ bin of the $j$-particle ToT distribution.  Equation~\ref{eq:seppower} is normalized to be between $0$ and $1$, where $0$ is only possible if the two distributions are identical.  Figure~\ref{fig:TwoParticleID:simple3} shows the dependence of the separation power on the number of bits and the tuning in the linear charge to ToT conversion scheme.  The separation power is largely insensitive to the tuning so long as the charge corresponding to the overflow bin is below the raise in the two-particle peak (in e.g. Fig.~\ref{fig:TwoParticleID:simple1}).  The right plot of Fig.~\ref{fig:TwoParticleID:simple3} zooms in on the region of $\text{ToT}_\text{MIP}$ between $0$ and $2\times\text{ToT}_\text{half}$.  As long as the number of bits is above 3-4, there is little sensitivity to the tuning in the relevant region $\text{ToT}_\text{MIP}\gtrsim\text{ToT}_\text{half}$.

\subsection{General case}
\label{sec:class:general}


Both the ATLAS~\cite{Aad:2014yva} and CMS~\cite{tidecms} collaborations use charge information to split pixel clusters in order to recover tracking efficiency inside high hit multiplicity environments.  In the case of the ATLAS algorithm, an artificial neural network based on the geometric distribution of the ToT inside a cluster is used to determine how many particles traversed the cluster.  This section uses a similar approach in order to build a realistic algorithm for entire cluster multiplicity classification.  

The neural network classifier uses the total cluster ToT and the geometric shape of the cluster.  A sample of one-particle clusters are produced by shooting a 10 GeV electron uniformly at random inside a given pixel.  The two-particle sample adds a second 10 GeV electron uniformly at random in the neighborhood of the first pixel.  For each charge to ToT setting (scheme, tuning, $n$ bits), the network is re-optimized.  Training is performed with the TMVA package~\cite{Hocker:2007ht} using one hidden layer with 10 nodes and the $\tanh$ activation.  

\begin{figure}[h!]
\centering
\includegraphics[width=0.6\textwidth]{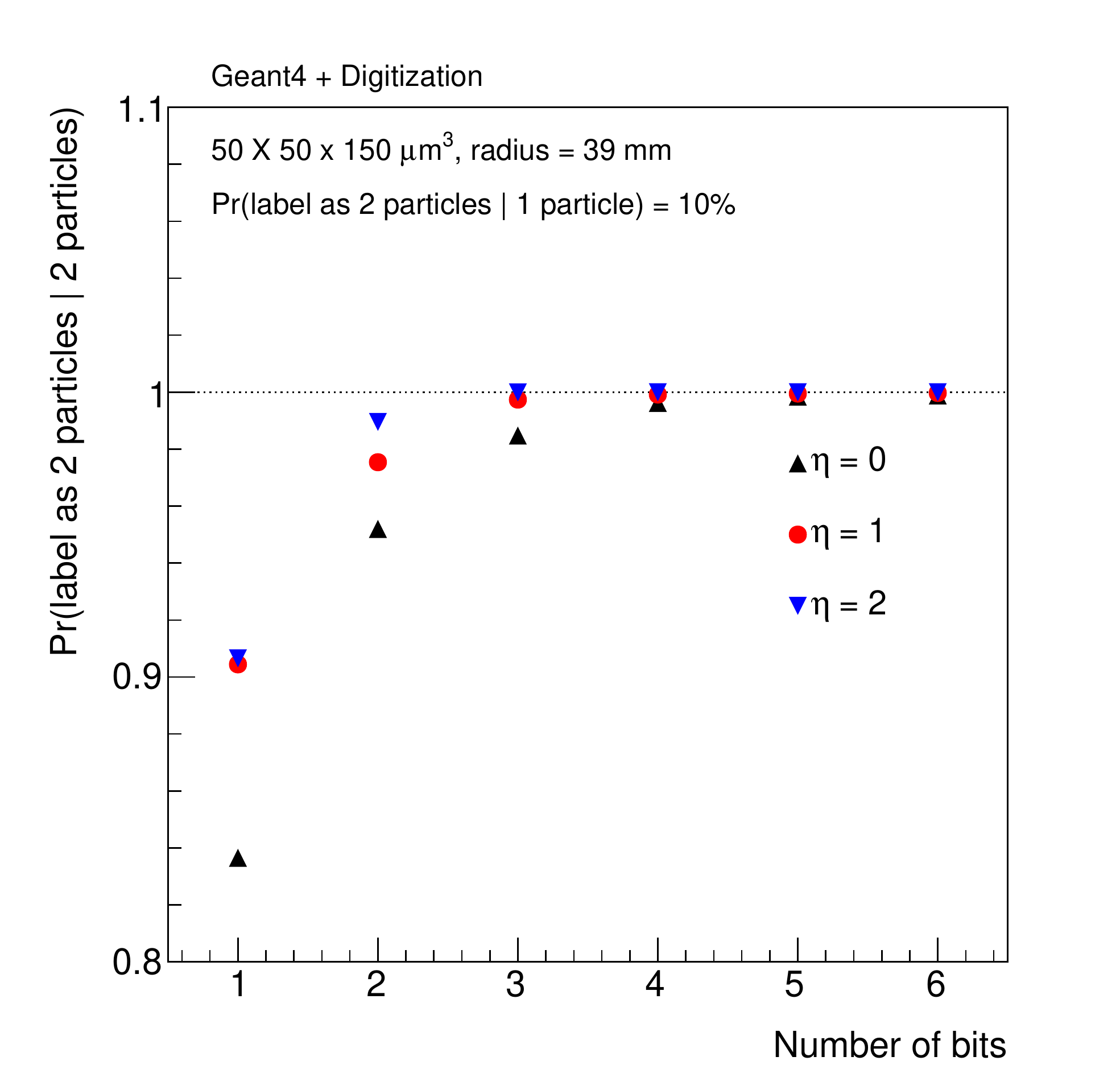}
\caption{The probability to correctly identify two-particle clusters (true positive) as a function of the number of bits in the linear charge to ToT conversion scheme, given that the probability to mis-identify a one-particle cluster as a two-particle cluster is $10\%$. }
\label{fig:TwoParticleID:NN1}
\end{figure}

There are many ways to quantify the performance of the neural network classifier.  The standard area under the curve (AUC) is not the most useful metric because one typically operates a fixed efficiency for identifying one-particle clusters correctly (since their prior probability is much higher).  Figure~\ref{fig:TwoParticleID:NN1} shows the probability to correctly identify a two-particle cluster (true positive rate) given that the probability to mis-classify a one-particle cluster (false positive) as a two-particle cluster is 10\%.  Since the charge is discretized into integers, it is not possible to achieve exactly 10\% false positive rate, so the nearest value is chosen.  For the long clusters at $\eta=2$, the performance saturates already at 3 bits; by 4-5 bits, saturation has occurred across the $\eta$ spectrum.

\clearpage

\section{Resolution}
\label{sec:resolution}

A charged particle track is specified by a series of fit parameters: $q/p, \theta,\phi,d_0,z_0$, where $q$ is the particle charge, $p$ is the momentum, $\theta$ is the polar angle, $\phi$ is the azimuthal angle, $d_0$ is the transverse impact parameter, and $z_0$ is the longitudinal impact parameter.  Precise knowledge of these parameters is critical for a wide variety of applications, ranging from the jet energy scale to $b$-tagging efficiency.   All of the track parameter resolutions derive from individual cluster position and length resolutions.  For example, the track $z_0$ is most sensitive to the $y$ position of the cluster in the innermost pixel layer.  This section explores the impact of charge digitization on the resolution of cluster properties, starting with single-particle clusters in the absence of $\delta$-rays.

\subsection{Single-particles without $\delta$-rays}

\begin{figure}[h!]
\centering
\includegraphics[width=0.6\textwidth]{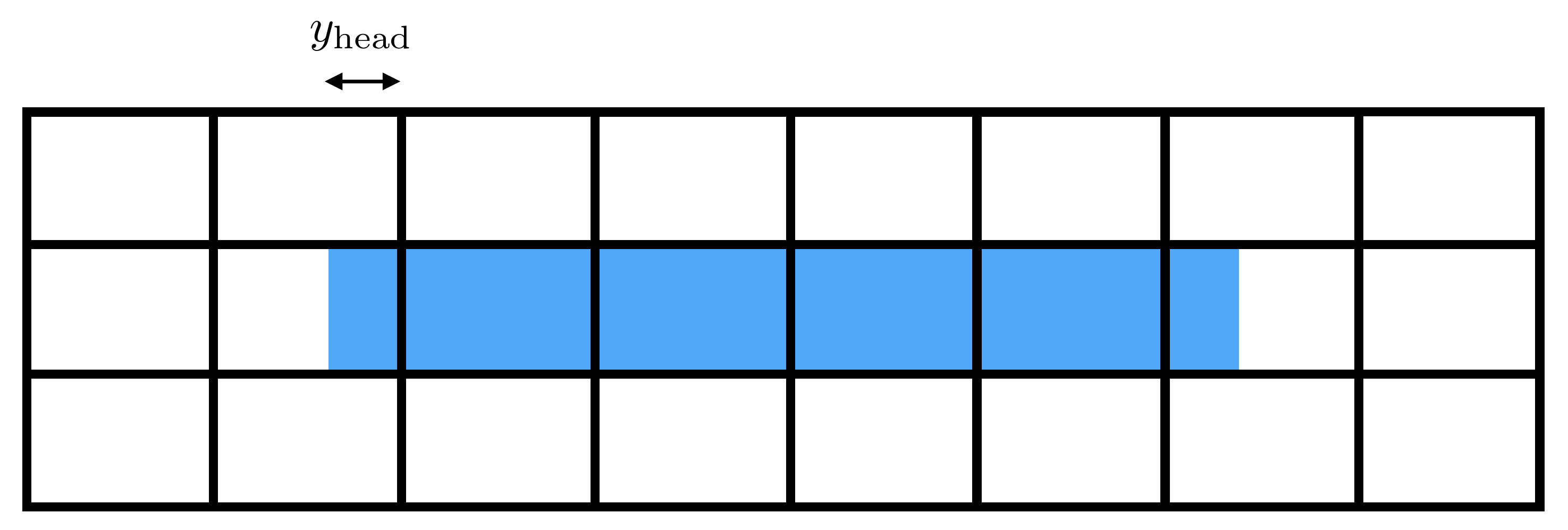}
\caption{A section of a pixel module as viewed from the top.  The filled in area represents the path traversed by a particle.  The location and length of the cluster are determined by the location of the head and tail.}
\label{fig:resolution:schematic}
\end{figure}

\begin{figure}[h!]
\centering
\includegraphics[width=0.5\textwidth]{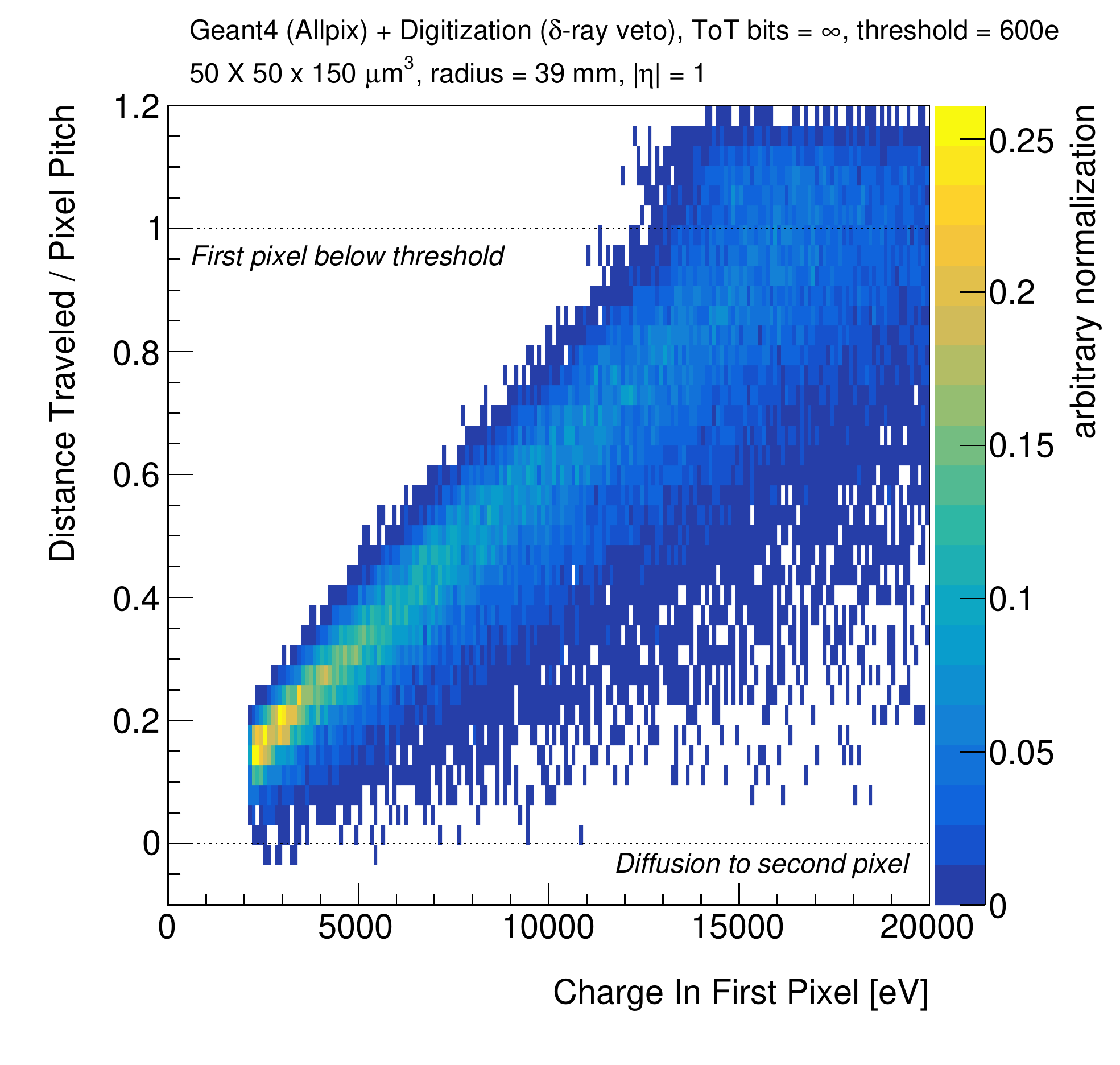}\includegraphics[width=0.5\textwidth]{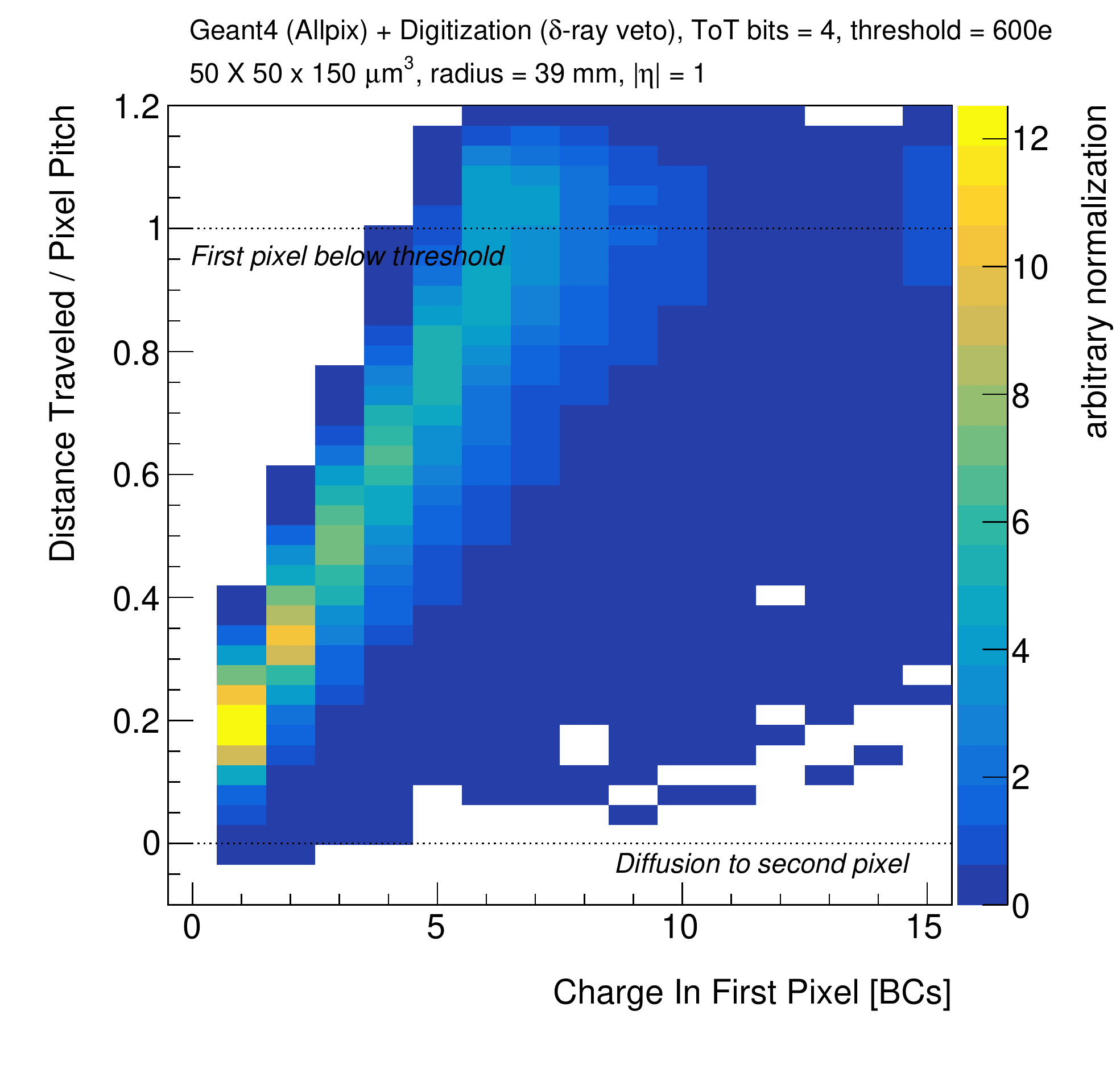}
\caption{The joint distribution of path length in the first pixel and the charge for arbitrary charge precision (left) and for 4 bits of ToT (right).  The variable plotted on the vertical axis is defined as $y_\text{$1^\text{st}$ pixel}-y_\text{enter}$, where $y_\text{$1^\text{st}$ pixel}$ is the $y$ position of the lower edge of the first pixel above threshold and $y_\text{enter}$ is the $y$ position where the particle enters the silicon (see Fig.~\ref{fig:resolution:classes}).  Distances are given in units of pixel pitch (50 $\mu$m).  The increased density at high ToT in the right plot is due to overflow.}
\label{fig:resolution:pathversuscharge}
\end{figure}

\begin{figure}[h!]
\centering
\includegraphics[width=0.5\textwidth]{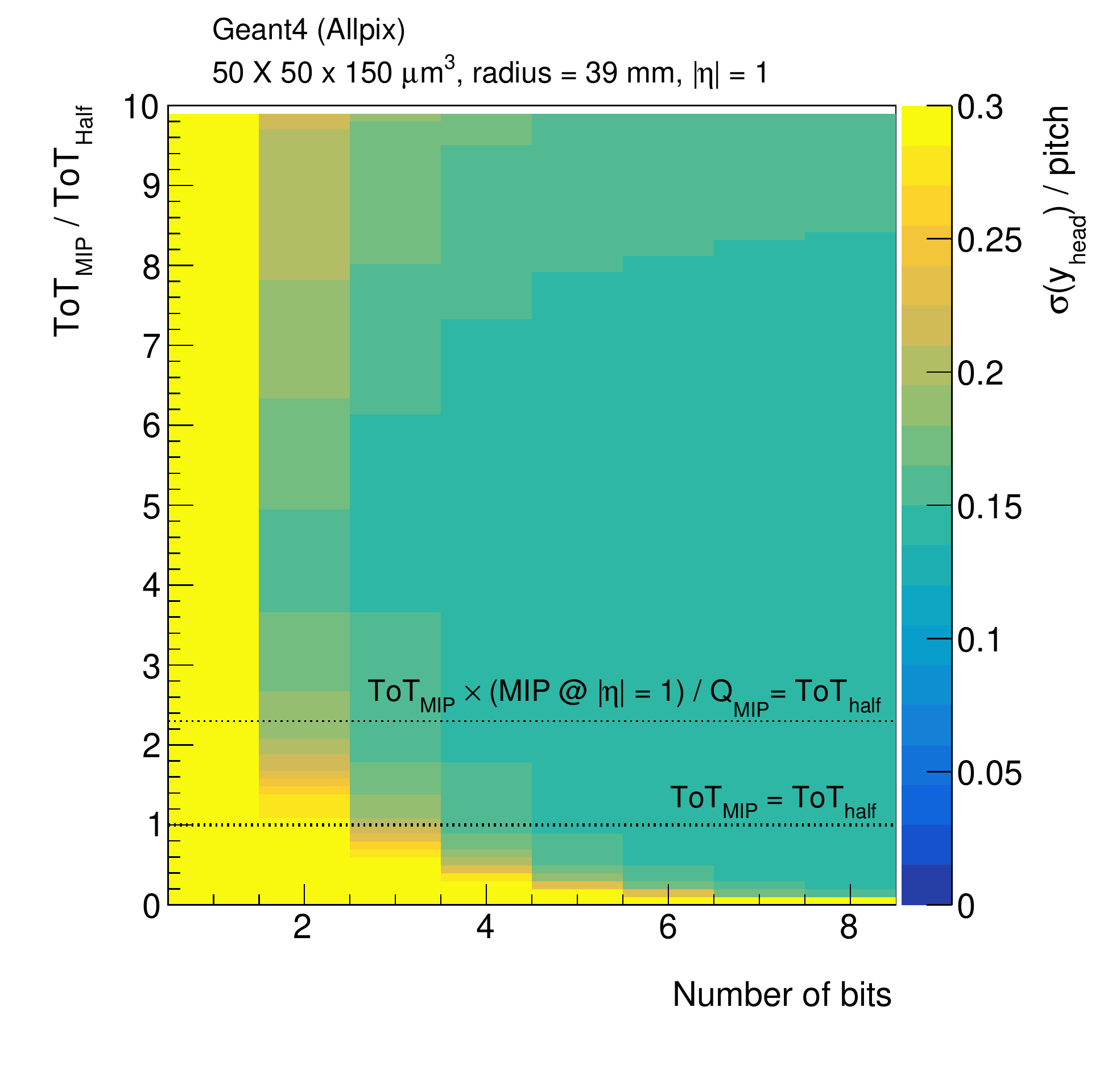}\includegraphics[width=0.5\textwidth]{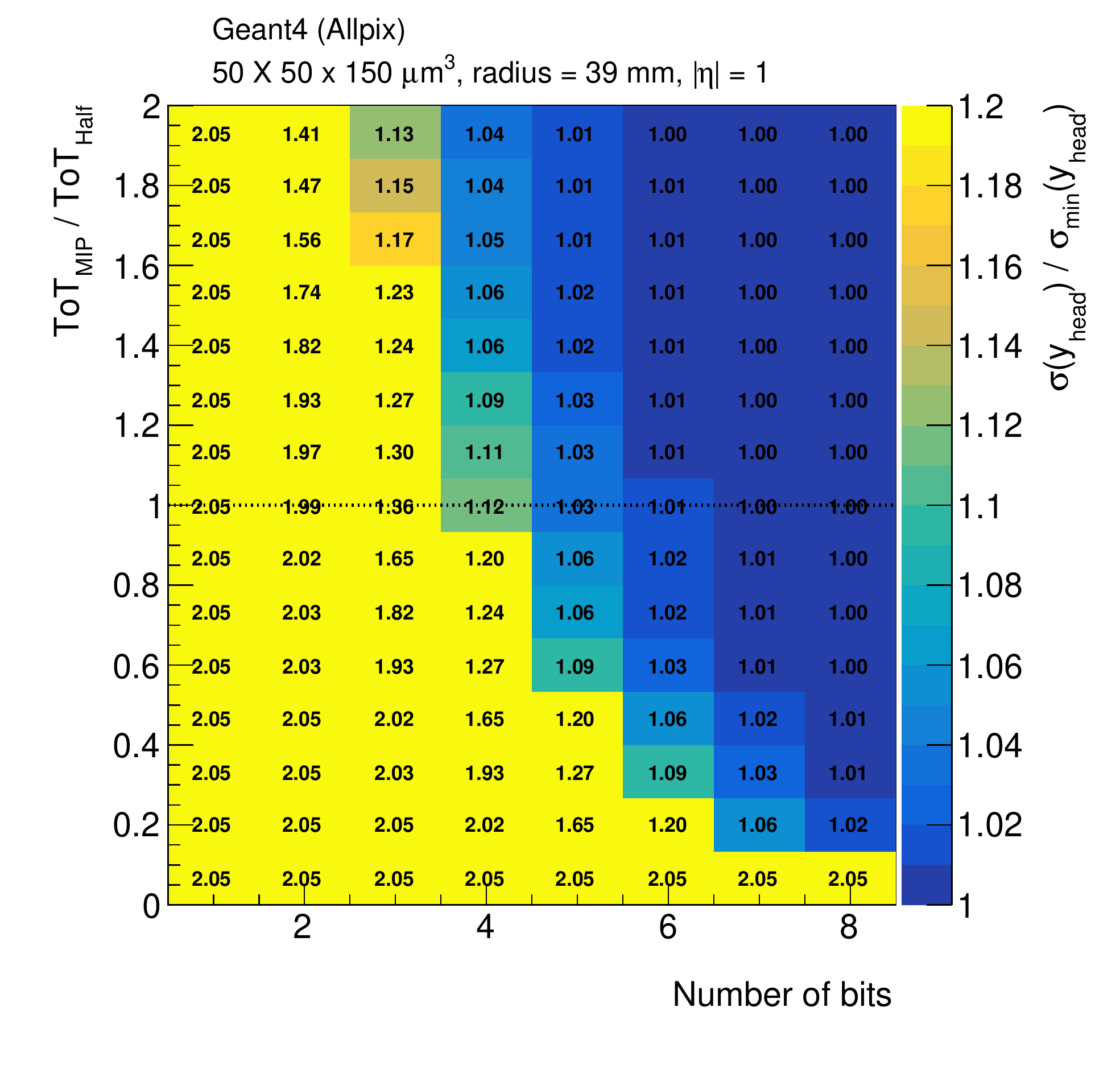}
\caption{Left: The optimal position standard deviation as a function of the number of ToT bits and the tuning used in the linear charge-to-ToT scheme.  To guide the eye, a horizontal line at one indicates the tune that corresponds to a MIP charge placed at half of the available ToT range.    To guide the eye, typical tuning values are indicated that correspond to $\text{ToT}_\text{MIP}=\text{ToT}_\text{half}$ and to where the actual MIP peak for $|\eta|=1$ correspond to half of the available ToT range.  Recall that the symbol $Q_\text{MIP}$ corresponds to 80 electrons/$\mu$m at perpendicular incidence. Right: A zoomed in version of the left plot where the resolution is divided by the minimum resolution, achieved with infinite ToT bits.}
\label{fig:resolution:scan}
\end{figure}

In the direction parallel to the beam, all of the information about a cluster's position and and length is contained in the position of the head and tail of the cluster: $y_\text{cluster}=\frac{1}{2}(y_\text{head}+y_\text{tail})$ and $L_\text{cluster}=y_\text{head}-y_\text{tail}$ (see Fig.~\ref{fig:resolution:schematic} for an illustration).  To a good approximation, the tail and head position resolution are identical and therefore $\sigma(y_\text{cluster})\approx\sigma_{y_\text{head}}/\sqrt{2}$ and $\sigma_{L_\text{cluster}}\approx\sqrt{2}\sigma_{y_\text{head}}$.  Without charge information (digital clustering), $\sigma_{y_\text{head}}\lesssim\text{pitch}/\sqrt{12}$.  The deposited charge scales with the path length and so charge information (analog clustering) can be used to improve the position resolution.  Figure~\ref{fig:resolution:pathversuscharge} shows the relationship between the path length and the deposited charge.  There is a clear increasing trend: when the particle only traverses a small fraction of the first pixel, the deposited charge is typically low.  Due to the large fluctuations in the charge about the mean, once a particle has traversed nearly an entire sensor, there is little information about its precise location inside the pixel.  Given a charge (or ToT), the procedure for picking a distance that minimizes the position standard deviation is to select the average distance for that charge.  Symbolically, $\hat{x}(Q) = \langle x|Q\rangle$, where $\hat{x}$ is the estimated position.  Figure~\ref{fig:resolution:scan} shows this optimized position resolution as a function of the number of bits of ToT and $\text{ToT}_\text{MIP}$ for the linear charge-to-ToT conversion scheme.  As expected, the resolution is poorer ($\sigma\sim\text{pitch}/\sqrt{12}$, though see Ref.~\cite{binaryreadout}) when the number of bits is small or the effective number of bits is small ($\text{ToT}_\text{MIP}\gg\text{ToT}_\text{half}$ or $\text{ToT}_\text{MIP}\ll\text{ToT}_\text{half}$).  For a tuning that puts the $|\eta|=1$ MIP peak at $\text{ToT}_\text{half}$, the performance saturates around 5 bits with $\sigma\sim 15\%\times\text{pitch}\sim8$ $\mu$m.

\begin{figure}[h!]
\centering
\includegraphics[width=0.95\textwidth]{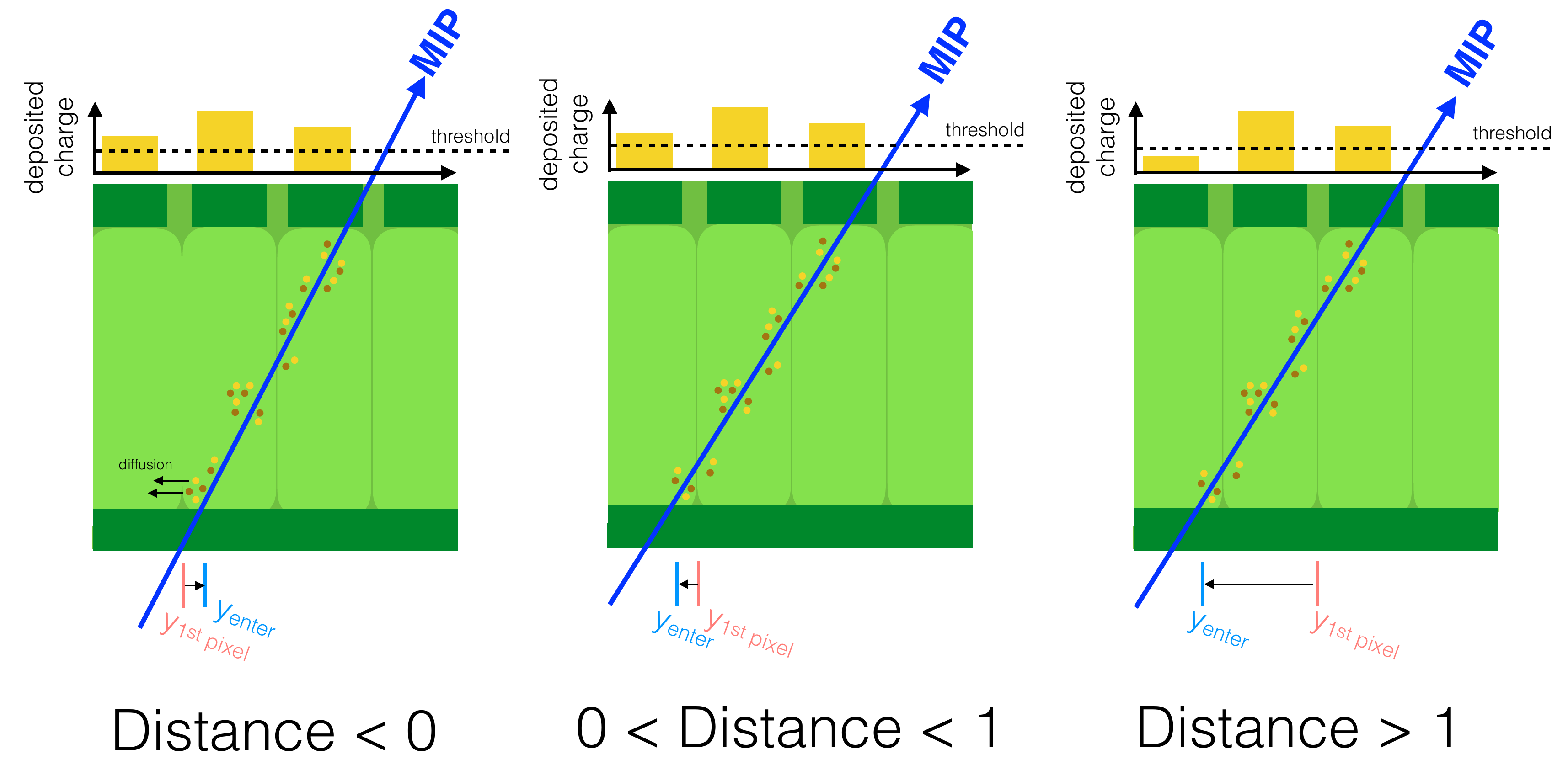}
\caption{A schematic diagram to show the calculation of the variable shown in the vertical axis (Distance) in Fig.~\ref{fig:resolution:pathversuscharge}.  The Distance can be negative (left) when the first pixel above threshold has a lower $y$ than where the particle entered the silicon (due to diffusion).  In addition, the Distance can be larger than one (right) when the first pixel the particle traverses below threshold.}
\label{fig:resolution:classes}
\end{figure}

\begin{figure}[h!]
\centering
\includegraphics[width=0.5\textwidth]{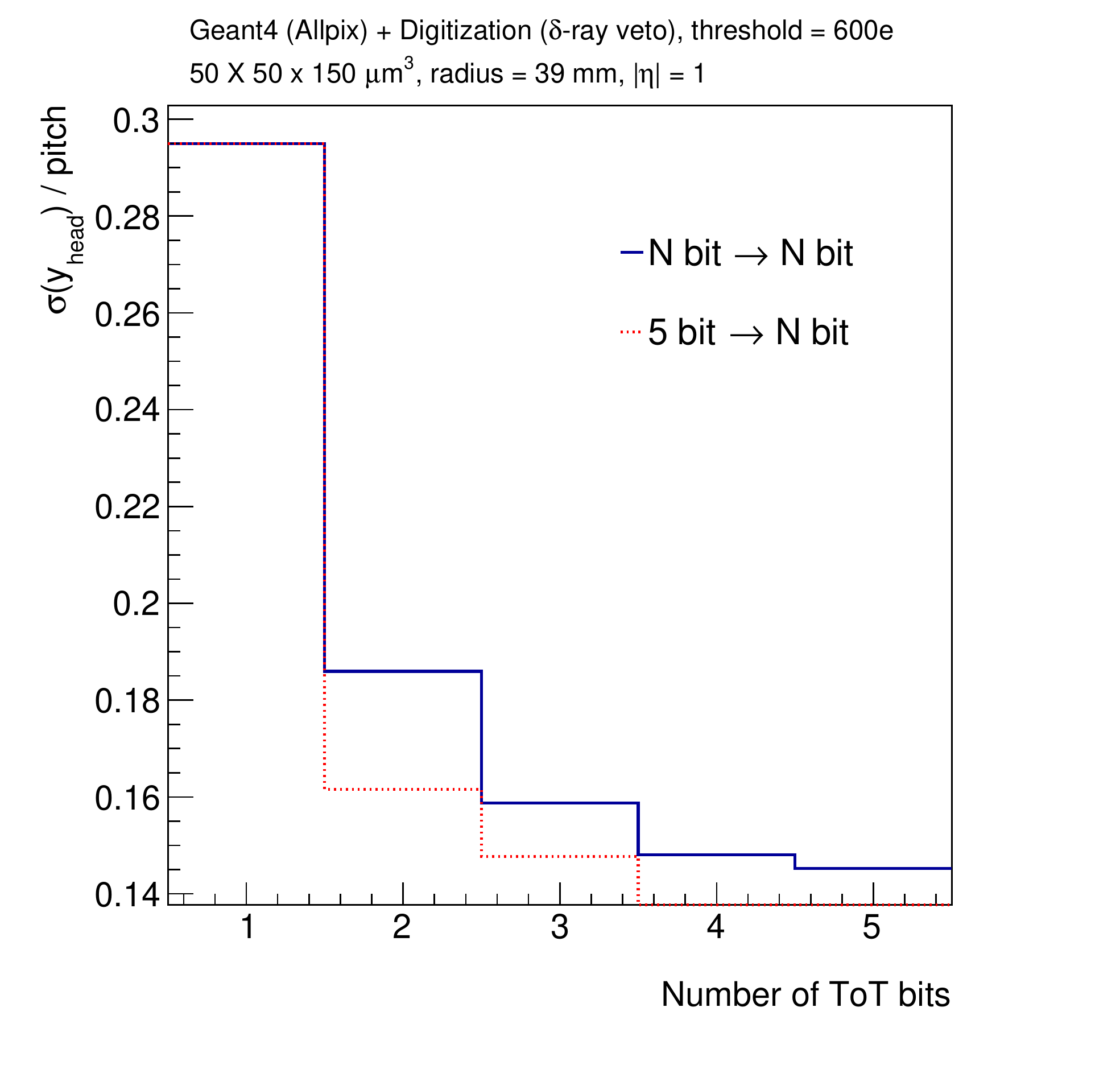}\includegraphics[width=0.5\textwidth]{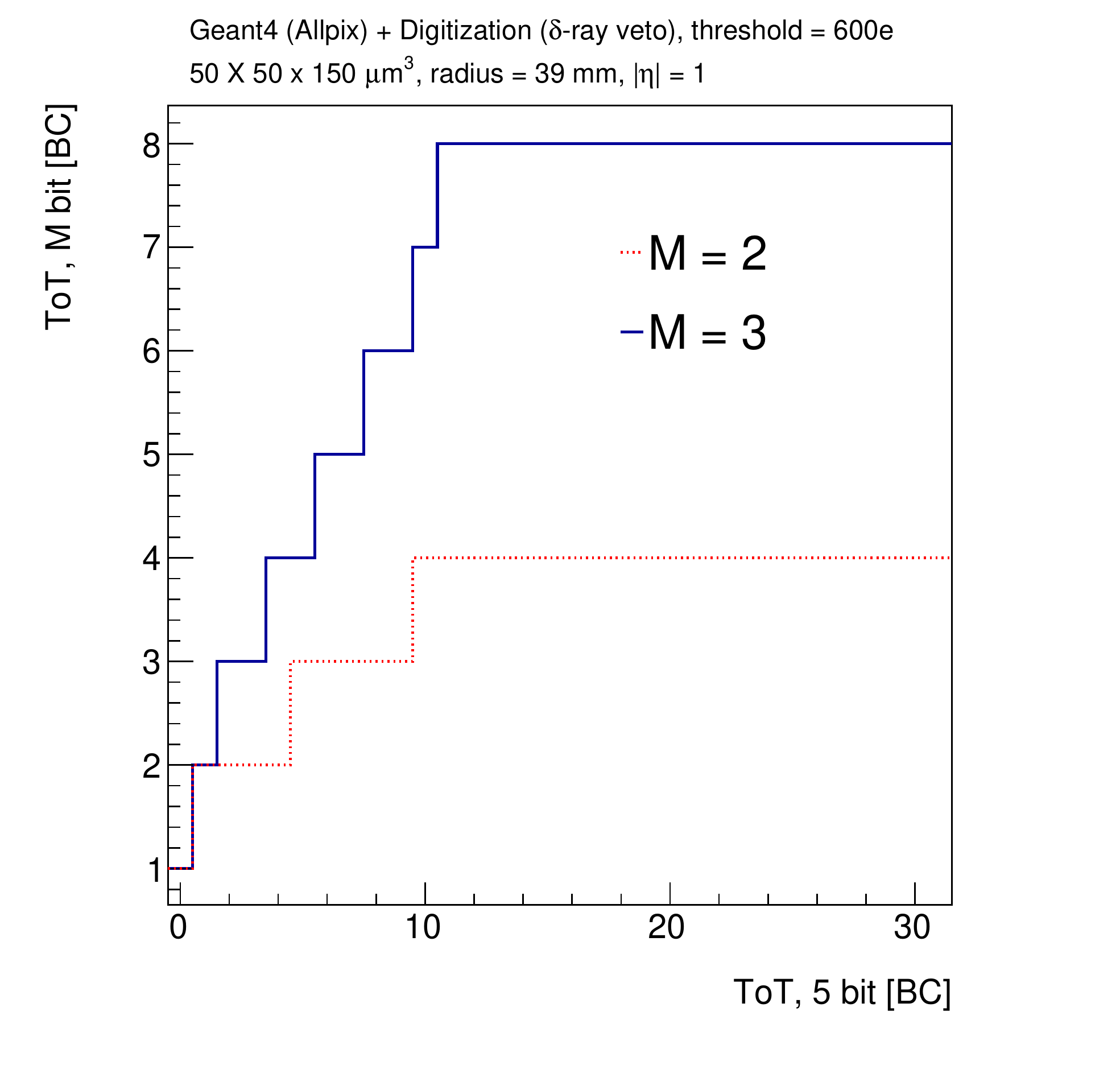}
\caption{Right: a comparison of the resolution when using a $n$ bit counter combined with a $n$ bit storage to the resolution when using a 5 bit counter combined with $n$ bits of storage.  Right: the optimal down-sampling functions for $5\rightarrow 3$ and $5\rightarrow 2$. }
\label{fig:resolution:nonlinear}
\end{figure}

One possible way to increase precision while reducing buffer space on chip is to use $m>n$ bits for counting the ToT and then quickly converting the $m$ bit ToT into $n$ bits for storage.  In general, there are $(2^m)^{2^n}$ possible functions that maps $2^m$ numbers onto $2^n$ numbers.  However, the number of actual possibilities is much smaller after removing obviously sub-optimal functions and symmetries from the space of functions.   Th actual number of unique mappings is ${2^m-1} \choose {2^n-1}$, as demonstrated in App.~\ref{sec:app}.  Therefore, a $m=5$ bit counter results in $4495$ maps with $n=2$, $3\times 10^6$ maps with $n=3$ and $3\times 10^8$ maps with $n=4$.  The left plot of Fig.~\ref{fig:resolution:nonlinear} shows the resolution for the optimal function for the various $m=5$ cases.  The resolution is nearly $\text{pitch}/\sqrt{12}$ when $n=1$ and reaches the full $m=n=5$ performance when $n=3$.  Unsurprisingly, the optimal $m$-bit word $\rightarrow$ $n$-bit word transformations $f$, indicated in  the right plot of Fig.~\ref{fig:resolution:nonlinear}, are of the form $f(i)=i$ for $i\leq 2^n-1$ and then $f(i)=2^n-1$ for larger values.  This is because the low charge region is most important for resolution, so as many bits as possible are allocated to that region.


\clearpage

\subsection{General single-particle clusters}
\label{sec:ressingleparts}

\begin{figure}[h!]
\centering
\includegraphics[width=0.5\textwidth]{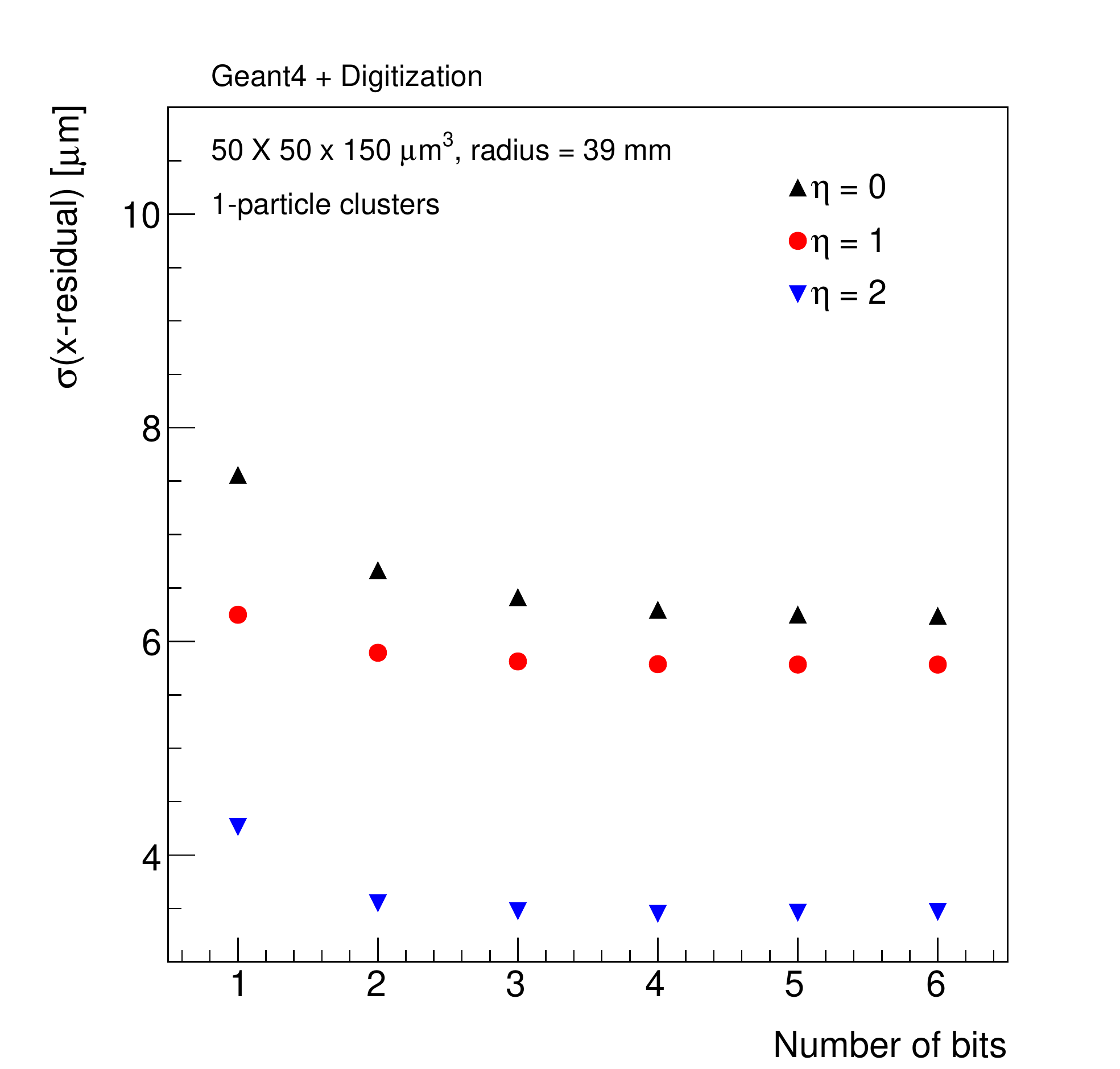}\includegraphics[width=0.5\textwidth]{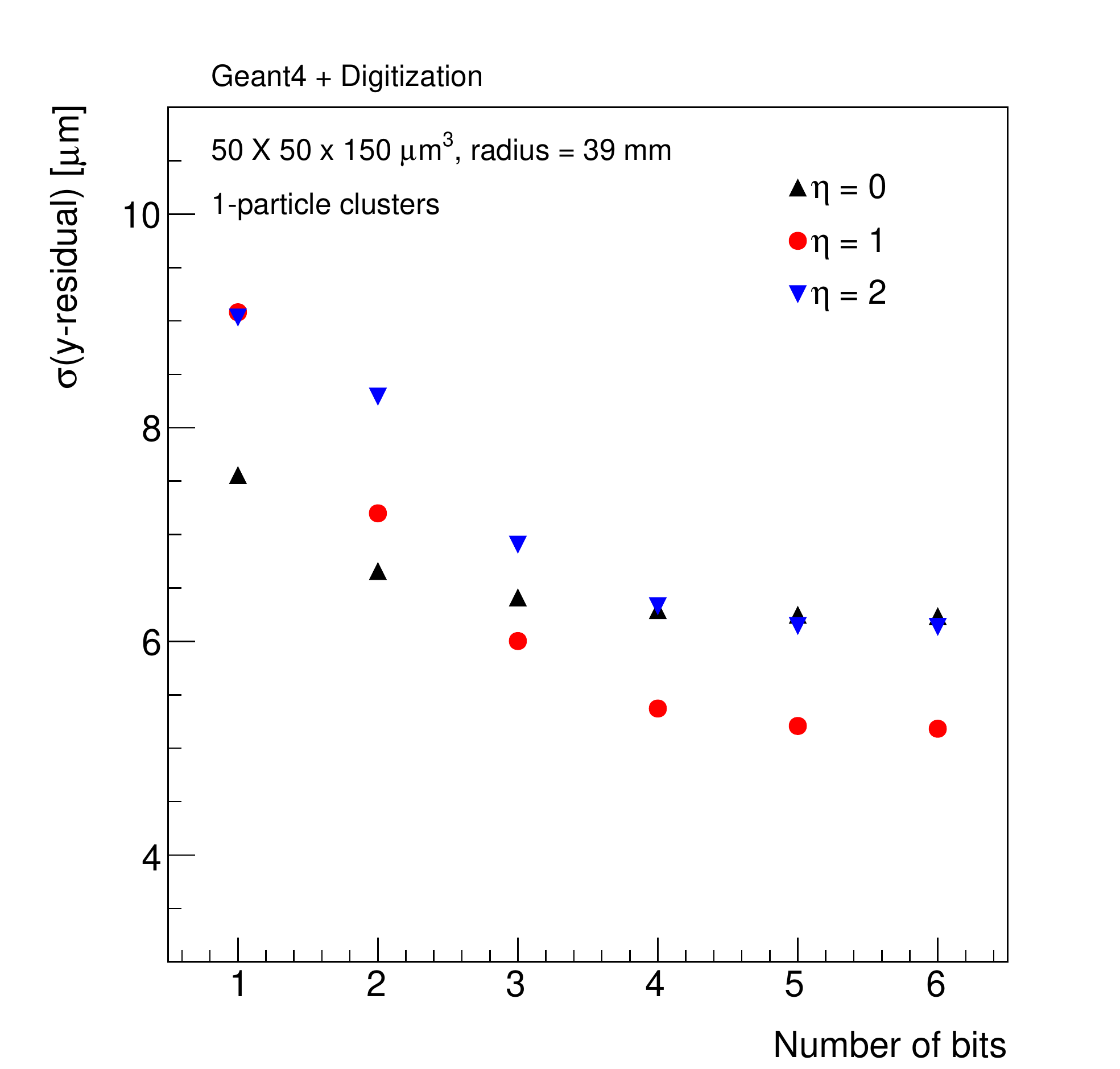}
\caption{The single-particle pixel cluster residual standard deviation in the $x$ ($y$) direction on the left (right) as a function of the number of bits of ToT using a neural-network-based regression for position estimation.  The NN is re-trained for each $\eta$ and each number of bits; this results in small statistical fluctuations that are approximately the same size as the markers.}
\label{fig:resolution:NN1}
\end{figure}

As in Sec.~\ref{sec:class:general}, the general case is approached with a neural-network based algorithm that identifies the position inside the pixels given the charge and shape information.  This regression problem is more complex than the classification one encountered in Sec.~\ref{sec:class:general} which is reflected in the neural network architecture: three hidden layers are now used, still with 10 nodes each and the $\tanh$ activation function.  For the long clusters at high $|\eta|$, nearly all of the information in the $x$ position is encoded by the ToT sum over the pixels in $y$ and vice versa for the $y$ position.  Therefore, the inputs to the NN are the pixel cluster row (or column) ToT sums as well as the number of hit pixels in each row or column.  The number of hit pixels is the most useful information when the position is in between two pixels (low charge regime).  It is important to explicitly provide this as an input to the NN because this information cannot be reconstructed from the row or column ToT sum when the number of bits is large\footnote{To see this, note that if there is only one bit, then the ToT sum is exactly the number of hit pixels; when the number of bits is infinite, there are a complete degeneracy between a little charge distributed on many pixels and a larger charge concentrated on a small number of pixels.}.   Figure~\ref{fig:resolution:NN1} shows the residual resolution for the $x$ and $y$ directions as a function of the number of bits.  As expected the $x$ and $y$ residual resolution is the same when $\eta=0$.  The residual resolution saturates around 2-3 bits in the $x$ direction and at about 4 bits in the $y$ direction.


\subsection{Multi-particle clusters}

\begin{figure}[h!]
\centering
\includegraphics[width=0.5\textwidth]{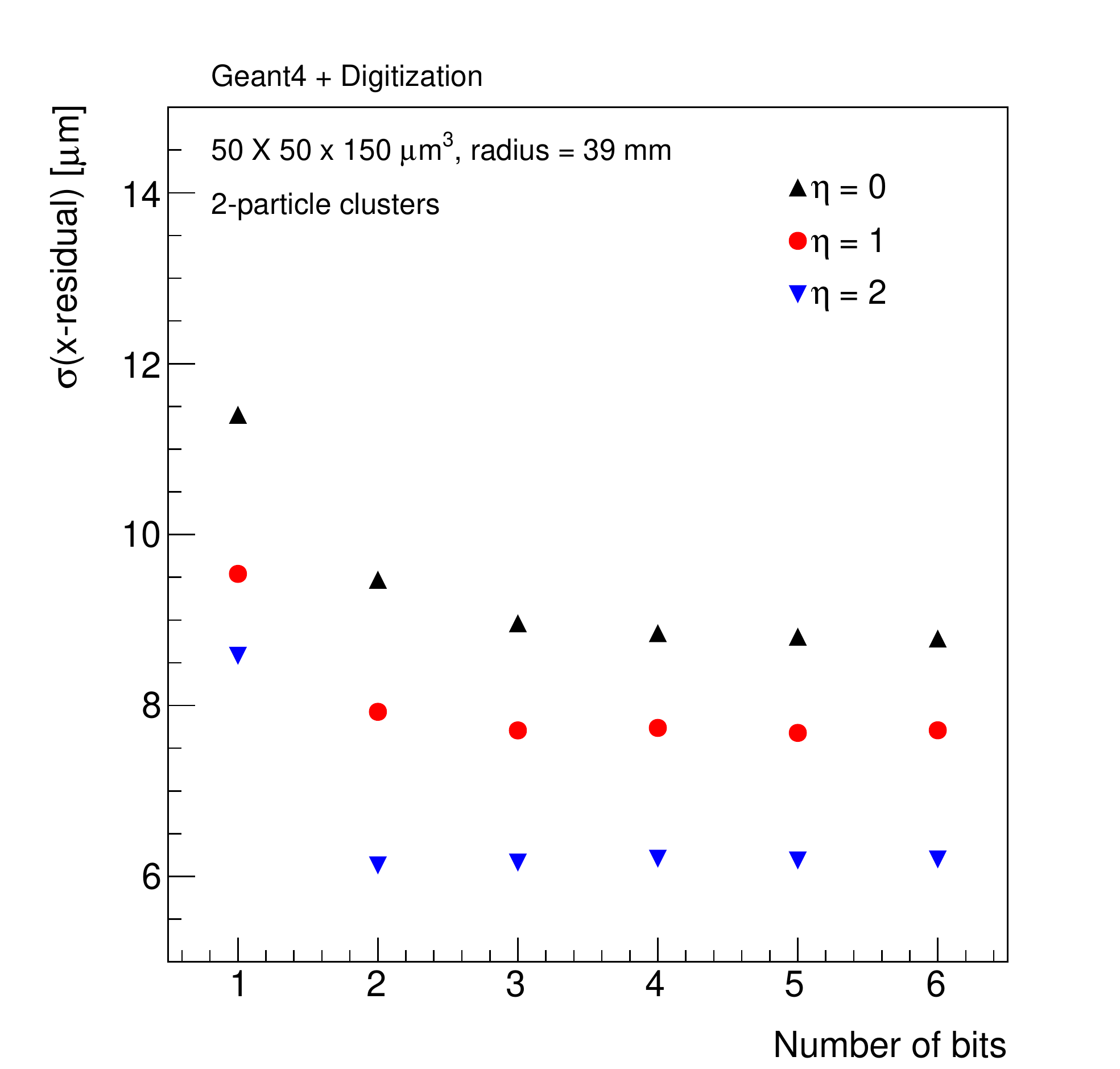}\includegraphics[width=0.5\textwidth]{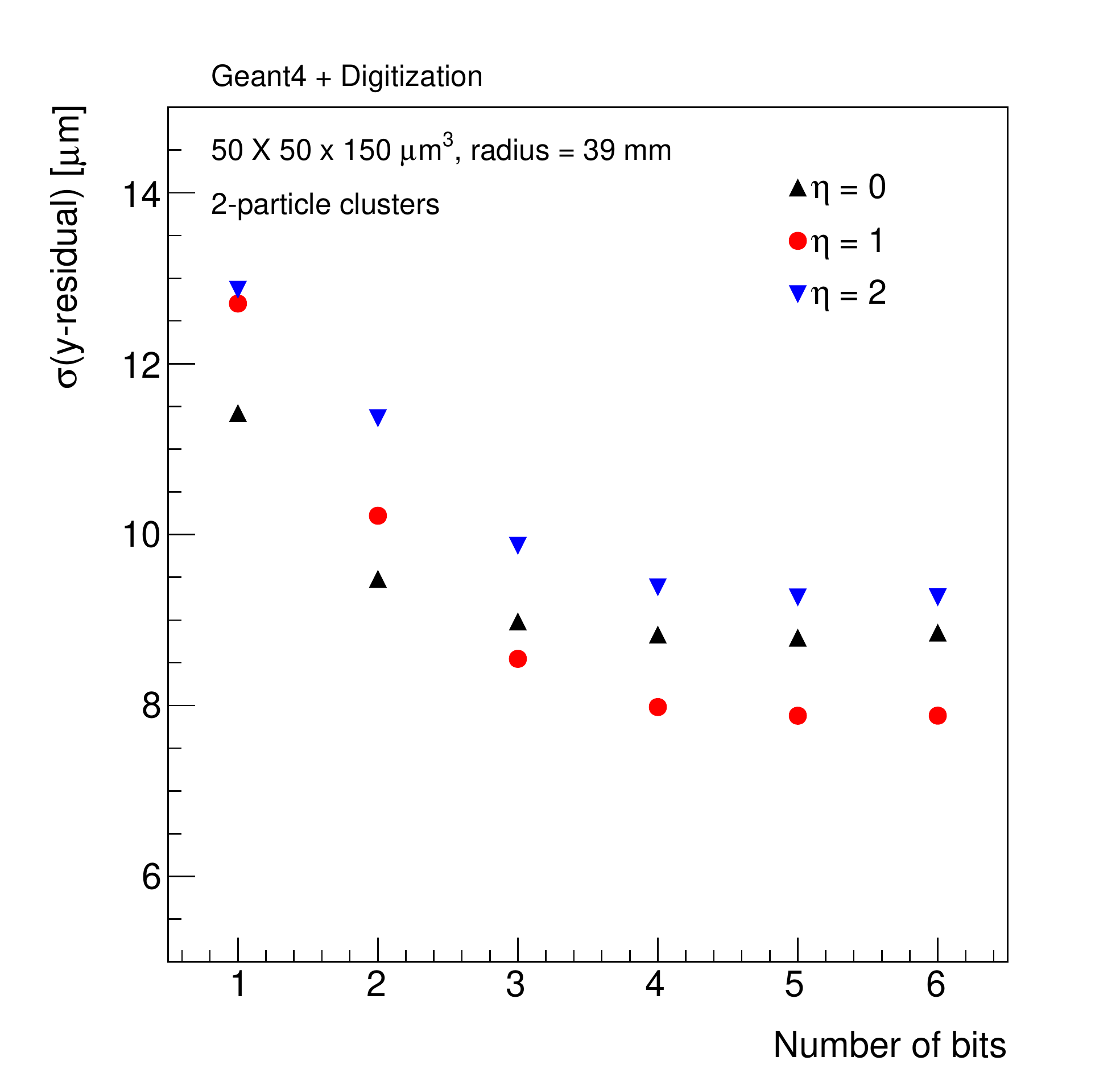}
\caption{The two-particle pixel cluster residual standard deviation as a function of the number of bits of ToT using a neural-network-based regression for position estimation.  The NN is re-trained for each $\eta$ and each number of bits; this results in small statistical fluctuations that are approximately the same size as the markers.}
\label{fig:resolution:NN2}
\end{figure}

For pixel clusters identified as originating from multiple charged particles, one can use the same NN setup as the previous section to predict the positions of all the particles.  Figure~\ref{fig:resolution:NN2} shows the residual resolution for the two-particle case as a function of the number of bits.  The same NN from Sec.~\ref{sec:ressingleparts} is used, but re-trained with two-particle cluster examples.  As this task is harder than the one-particle case, the overall residual resolution in Fig.~\ref{fig:resolution:NN2} is worse than in Fig.~\ref{fig:resolution:NN1}.  However, the convergence with the number of bits is similar - the saturation happens around 2-3 bits for $x$ and around 3-4 for $y$.

\clearpage

\section{Particle Identification}
\label{sec:pid}

In addition to distinguishing the number of particles traversing a cluster, the ToT can be used to identify the particle type.  The average charge deposited in silicon follows the Bethe equation, which depends on the $\beta\gamma=p/m$ of the incident particle.  Minimum ionizing particles have an energy loss in the broad minimum of the Bethe curve near $\beta\gamma\sim 1$.  However, there are many uses for identifying particles whose energy loss is well above $Q_\text{MIP}$.  For example, low energy $\delta$-rays are more ionizing than the primary MIP and have a characteristic charge profile due to the Bragg peak at the end of their trajectory when they are stopped.    Identifying $\delta$-rays is very important for improving the cluster position resolution.  Section~\ref{sec:deltas} explores dE/dx-based $\delta$-ray identification.  In addition, there is an entire class of searches at the LHC for physics beyond the Standard Model that exploit the charge measurement of pixel detectors.  This includes searches for massive long lived unstable particles (LLPs), highly ionizing particles (HIPs), and heavy stable charged particles (HSCPs)~\cite{Aaboud:2016dgf,Khachatryan:2016sfv,ATLAS:2014fka,Chatrchyan:2013oca,Aad:2011yf,Aad:2012pra,Aad:2013pqd,Chatrchyan:2012sp,Khachatryan:2011ts}.  Section~\ref{sec:llp} examines the sensitivity to the number of bits for one particular benchmark LLP search.  Unlike the resolution section (Sec.~\ref{sec:resolution}), Sec,~\ref{sec:deltas} and~\ref{sec:llp} are both examining the sensitivity to very large amounts of charge, often more than the two MIP case discussed in Sec.~\ref{sec:classification}.

\subsection{$\delta$-rays}
\label{sec:deltas}

\begin{figure}[h!]
\centering
\includegraphics[height=0.4\textwidth]{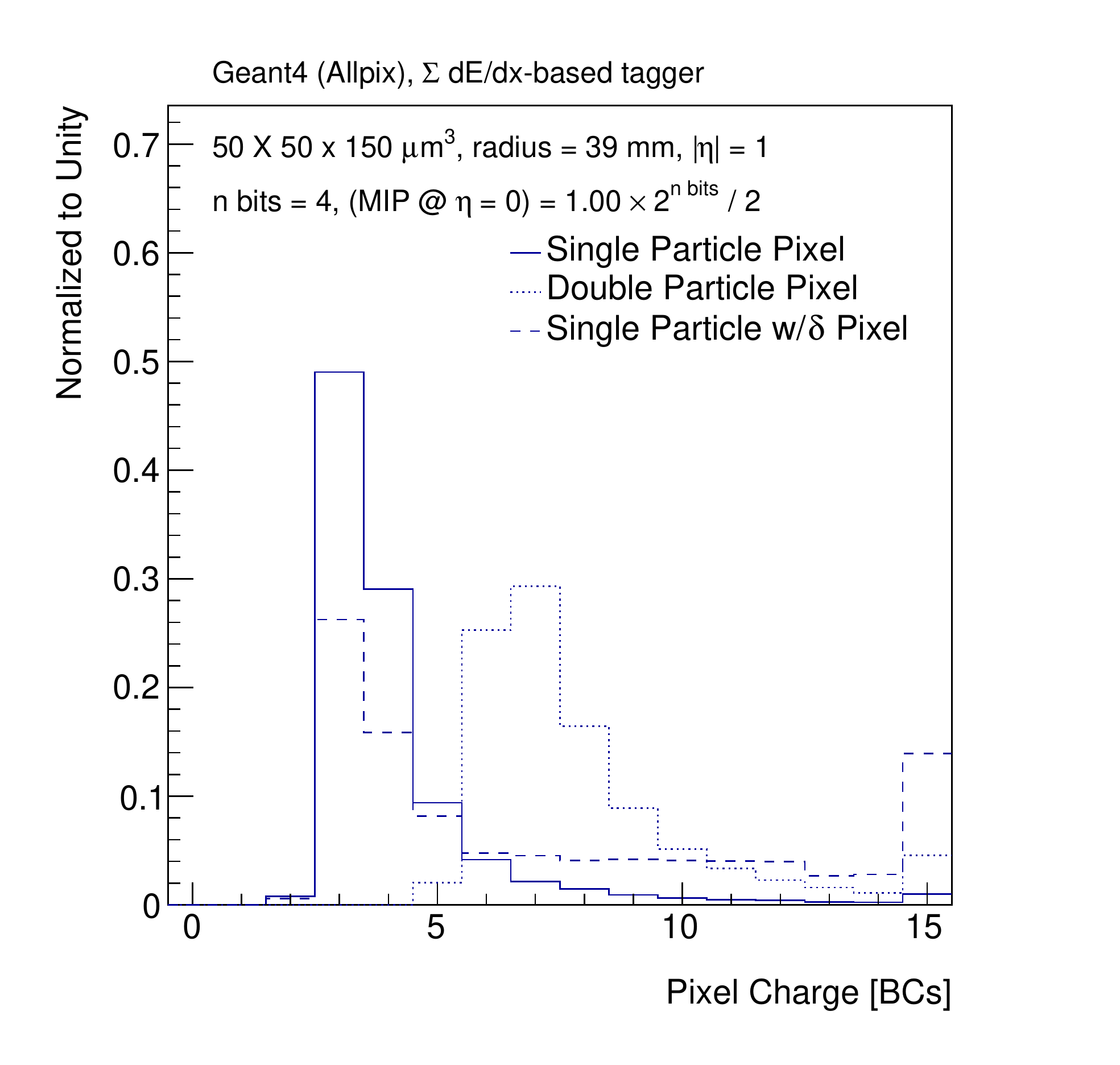}\includegraphics[height=0.4\textwidth]{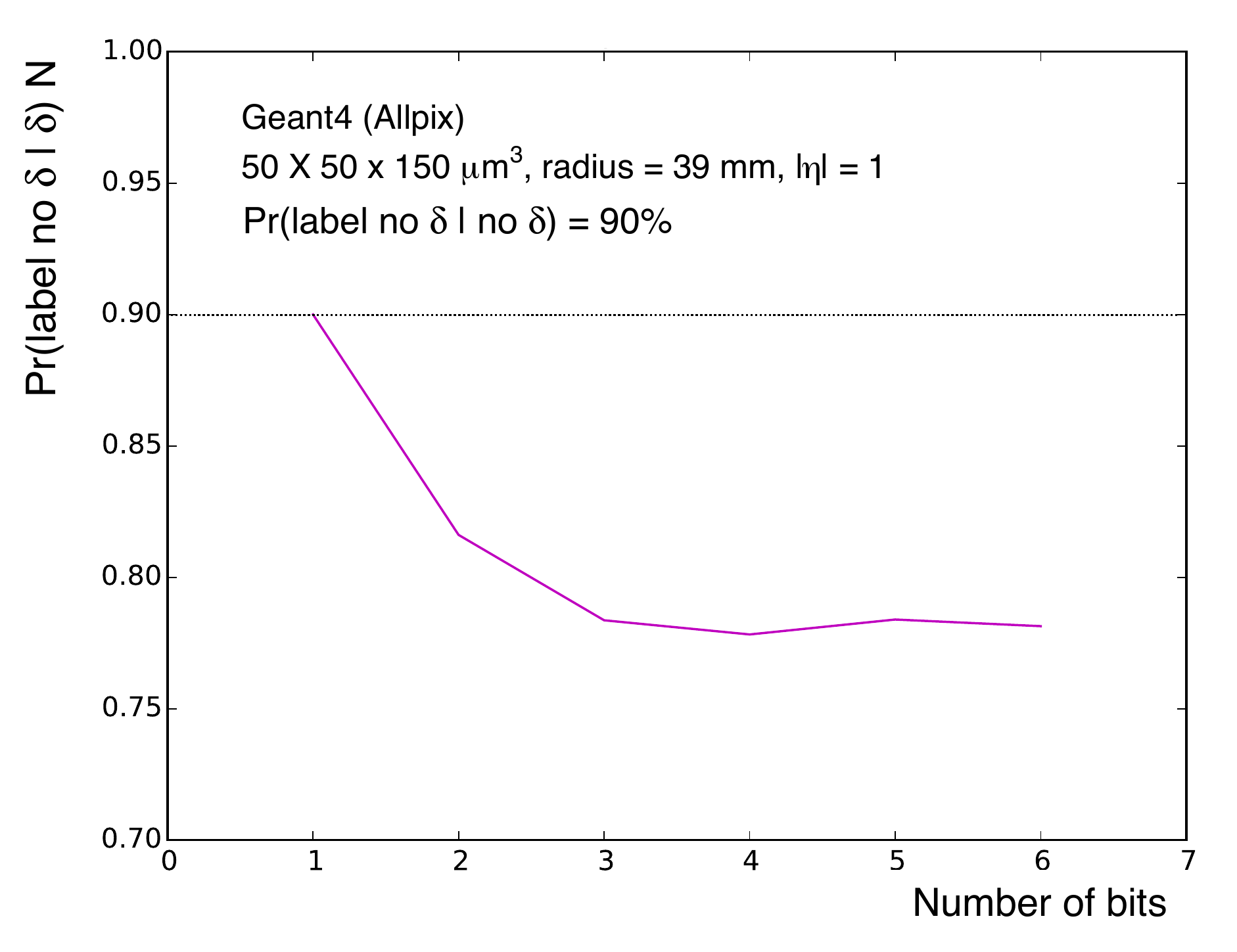}
\caption{Left: the ToT distribution for single pixels in a cluster (excluding head and tail pixels) resulting from a single MIP, two collinear MIPs, and a single MIP with a $\delta$-ray.  Right: the probability to mis-label a pixel due to a $\delta$-ray as a function of the number of ToT bits in the linear charge-to-ToT conversion scheme.  The probability to correctly classify a pixel without a $\delta$-ray is fixed at 90\%.  Linear interpolation is used in the ROC curve to achieve this 90\%, which results in the small apparent increase at high number of bits.}
\label{fig:deltarays}
\end{figure}

Knock-out electrons ($\delta$-rays) can travel a significant distance (many pixel lengths) before being stopped and their trajectory is largely uncorrelated with the primary ionizing particle.  Therefore, a naive cluster position reconstruction algorithms would result in a biased position when $\delta$-rays are present.  The energy profile of a $\delta$-ray changes  along its path length, with a significant raise just before it stops (Bragg peak).  Even though the dE/dx of a long $\delta$-ray can be MIP-like, the shapes of such $\delta$-ray tracks are easy to identify.  In contrast, $\delta$-rays that traverse only a short distance cannot be identified by their shape alone.  This section quantifies the classification performance for single pixel $\delta$-rays.  Since they are traveling slowly through the single pixel, such $\delta$-rays deposit more energy than a single MIP.  The left plot of Fig.~\ref{fig:deltarays} shows the ToT distribution for single pixels resulting from single MIPs, double MIPs, and $\delta$-rays. The latter has a much longer tail than a MIP and so identifying clusters with an anomalously high charge can be used to classify pixels that were likely due to $\delta$-rays.  The right plot of Fig.~\ref{fig:deltarays} shows the probability to mis-label a pixel due to a $\delta$-ray as a function of the number of ToT bits available to digitize the charge.  Since the probability for a pixel to be from a $\delta$-ray is small ($\sim 10\%$), it is important to maintain a low mis-classification rate of non-$\delta$-ray pixels.  The probability to mis-label a $\delta$-ray pixel as a single MIP pixel is 10\% with only binary readout and then saturates near $5\%$ with $\gtrsim 3$ bits of ToT. 

\subsection{Exotic New Particles}
\label{sec:llp}


A plethora of models for physics beyond the Standard Model predict new particles with a large charge and/or low $\beta\gamma$ that are also relatively long-lived ($c\tau\gtrsim \mathcal{O}(\text{mm})$).  Unlike many other searches for new particles, these theories present the exciting opportunity to directly detect the new particles from their interactions with the detector.  In particular, highly charged and/or slow-moving particles are expected to be more ionizing than MIPs and the combination of their velocity and ionization energy (along with an electric charge hypothesis) provides an estimate for the new particle mass. The most probable value of dE/dx can be approximated from a truncated mean over multiple pixel layers, in which the hit with the highest charge is dropped from the average.  By measuring a particle's momentum $p$ from the curvature of the reconstructed track and the track's truncated mean dE/dx, one can calculate a particle-by-particle mass with a parameterization of the Bethe formula that has been calibrated on a sample of MIPs.

\begin{figure}[h!]
\centering
\includegraphics[height=0.4\textwidth]{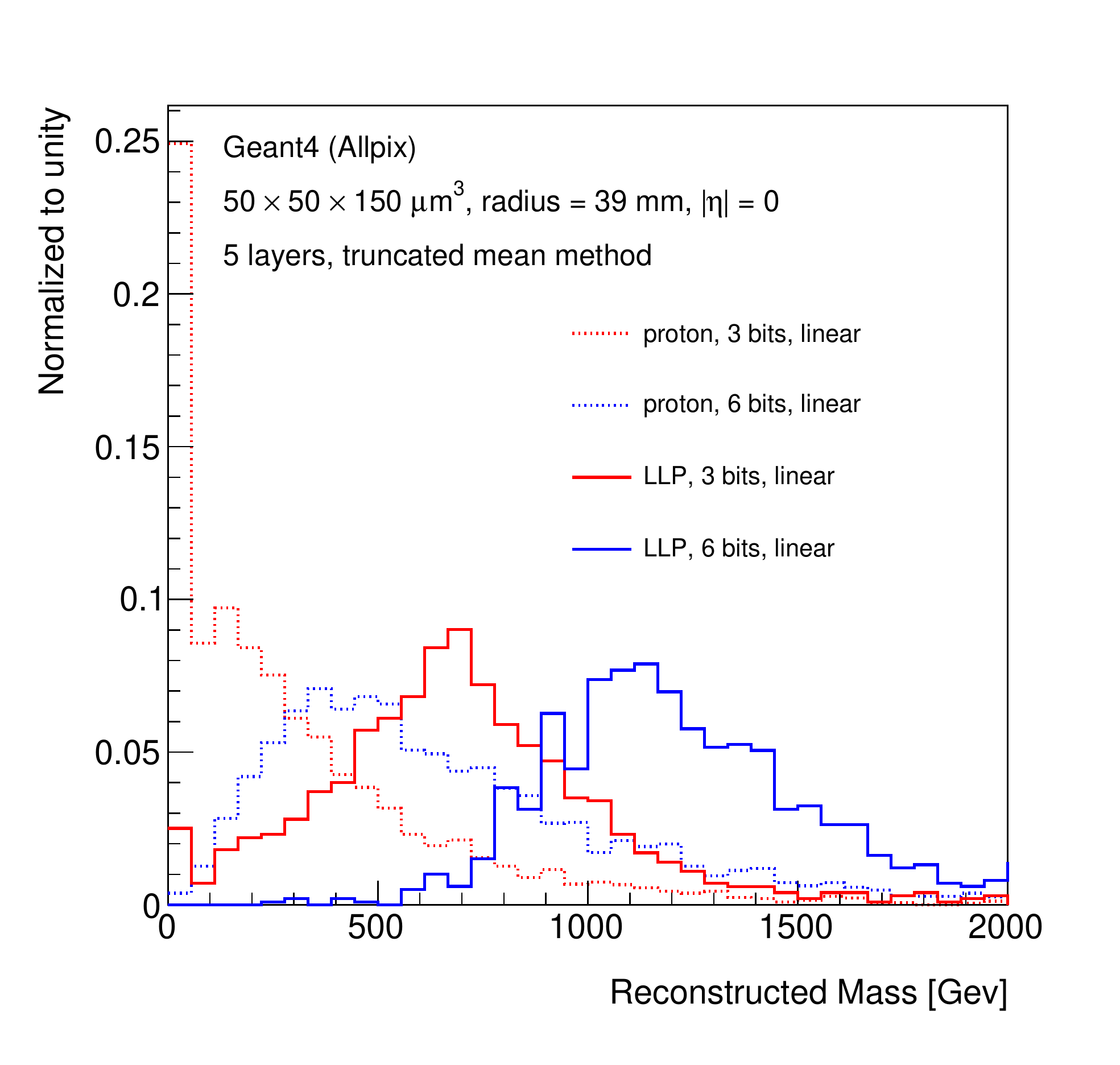}
\includegraphics[height=0.4\textwidth]{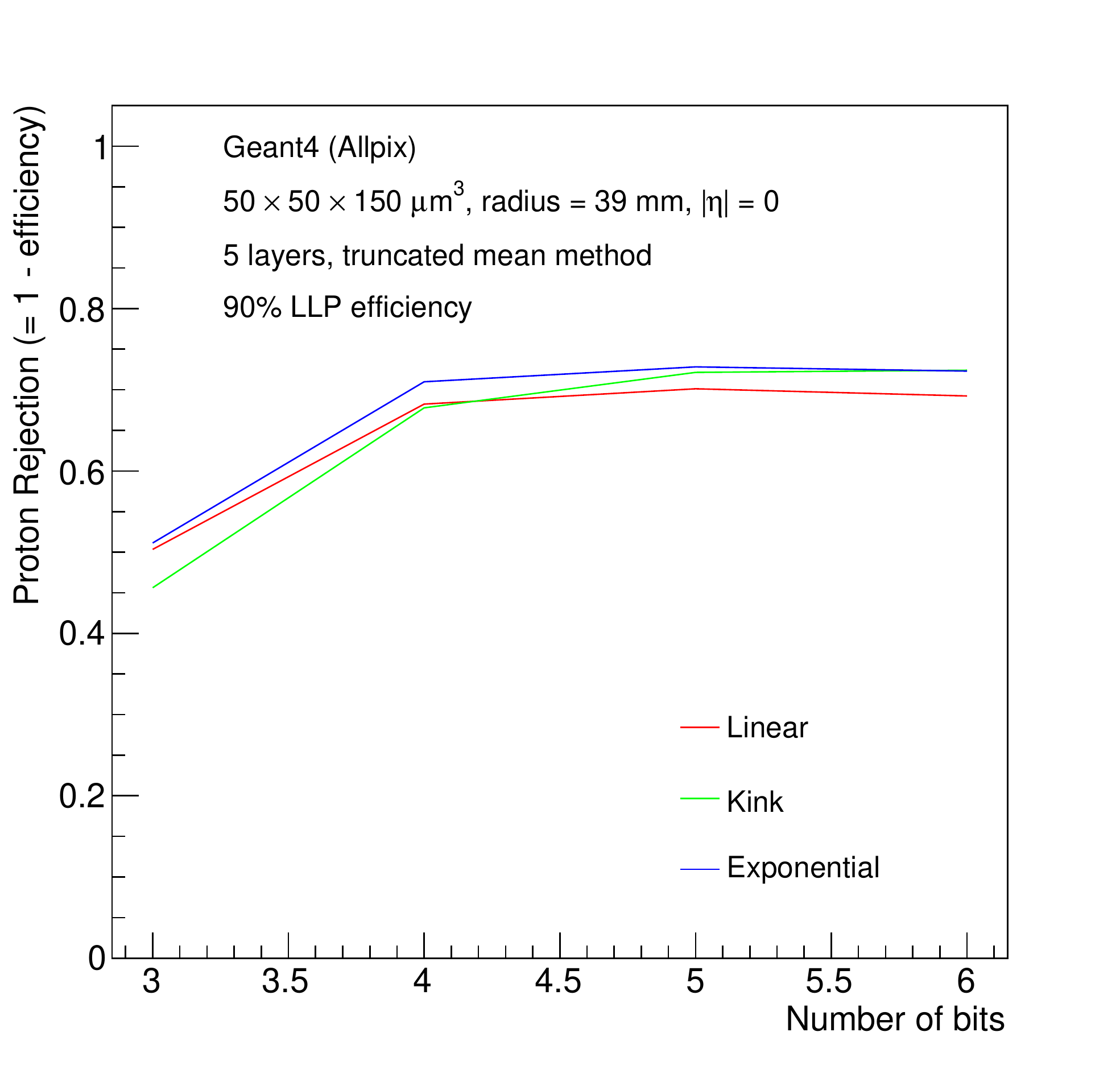}
\includegraphics[height=0.4\textwidth]{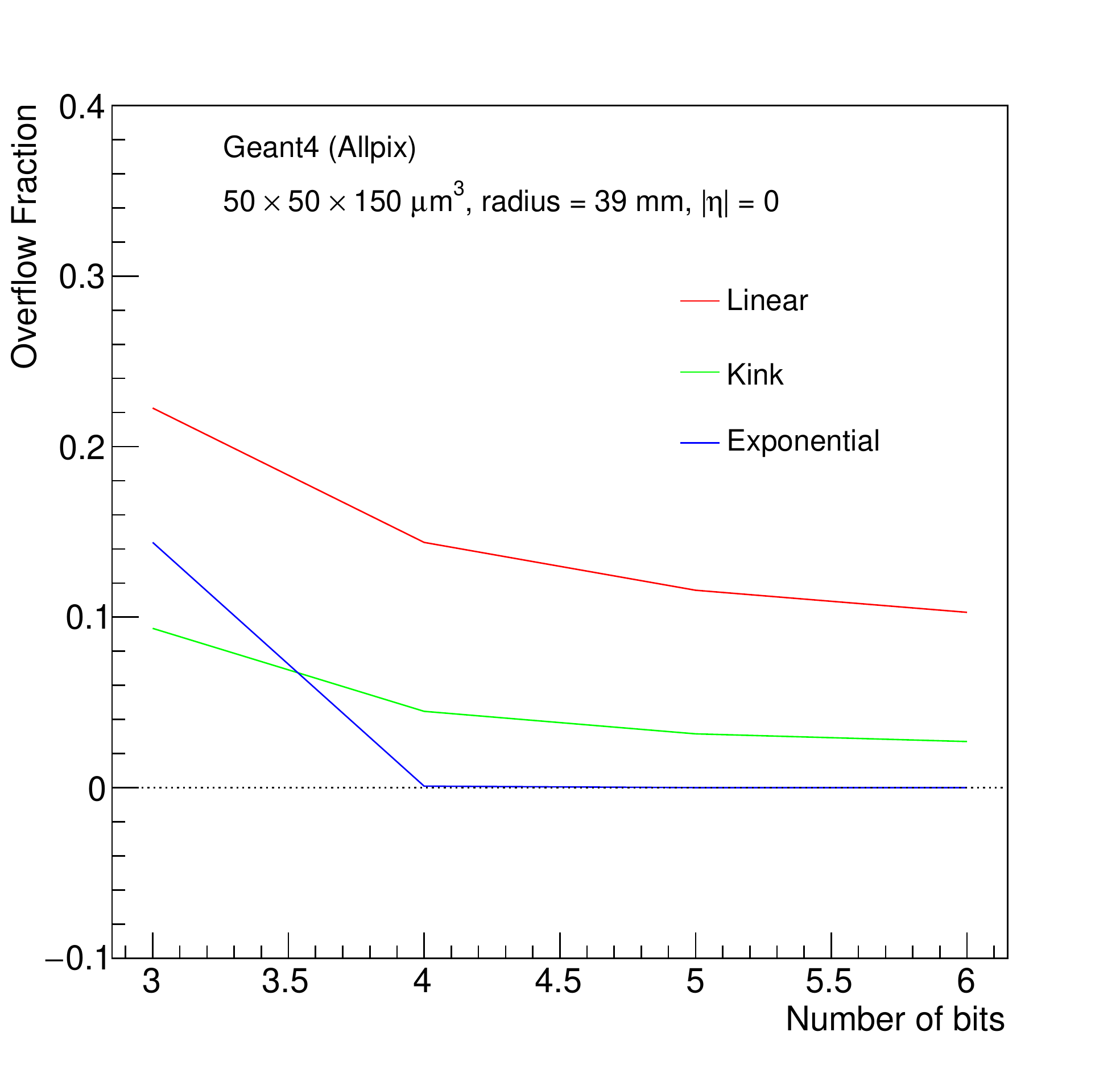}
\caption{Top left: the reconstructed mass of a MIP proton (dashed) and a 1.5 TeV slow moving R-hadron.  In a pixel detector, the  hadronic interactions of the R-hadron are negligible and the electromagnetic interactions are determined by its electric charge, mass, and velocity.  The mass is reconstructed using a five layer detector, dropping the most ionizing cluster. Top right: MIP proton rejection (=1-MIP proton efficiency) as a function of the number of ToT bits for a 90\% efficiency to select a 1.5 TeV LLP.  Bottom: The fraction of pixels in the ToT overflow for the various schemes.}
\label{fig:LLP}
\end{figure}

This section explores the classification of a 1.5 TeV long-lived R-hadron with electric charge $\pm 1e$.  An R-hadron is the hypothetical bound state of a long-lived (colored) gluino with standard model quarks~\cite{Farrar:1978xj}. With large momentum but $\beta<1$ due to their large mass, one can search for R-hadrons by selecting tracks with large $p$ and high dE/dx.  The reconstructed mass is a useful observable for discriminating signal from background. The background to searches for R-hadrons is mostly due to the long tail of the MIP dE/dx distribution. To approximate this background, the dE/dx distribution for a sample of protons is generated. In this study, the protons? momenta are assumed to follow the same distribution as that of the R-hadrons, in order to isolate the separation power of dE/dx alone.  The top left plot of Fig.~\ref{fig:LLP} shows the reconstructed mass distribution for MIPs and a 1.5 TeV LLP for two different numbers of ToT bits.  A large mass can be reconstructed for the MIPs due to their long dE/dx tail.  As expected, the mass resolution is better for 6 bits of ToT compared with 3 bits.  The reconstructed mass is used to distinguish the MIPs from LLPs.  This is quantified in the top right plot of Fig.~\ref{fig:LLP}.   The classification power of the reconstructed mass shows little dependence on the number of bits above 3 and also the advantage of the schemes that increase the charge range above the MIP peak are only slightly better ($\lesssim 5\%$) than the standard linear scheme.   This is true even though the fraction of clusters with a pixel in the ToT overflow is much less for the exponential scheme compared with the linear scheme (bottom plot of Fig.~\ref{fig:LLP}).  Therefore, $\gtrsim 4$ bits is likely sufficient for classification (at least for this benchmark model) while a precise measurement of the mass would require more bits.  

\section{Conclusions and Future Outlook}
\label{sec:concl}

In this paper, we have studied the impact of digitization on the use of ionization information from a pixel detector.  There are a variety of applications that can benefit from pixel charge information and therefore we have investigated four key areas: measurement efficiency, multiplicity classification, position resolution, and particle identification.  As the LHC experiments design their pixel readout chip for the HL-LHC, now is the time to decide what are the requirements to maximally exploit the data for physics analysis.  The studies presented in this paper suggest that a linear charge to ToT scheme is likely sufficient for a broad range of applications, however, there are possible gains by using a non-linear scheme.  Furthermore,  the performance seems to saturate around $4$-$5$ bits of ToT.  Using these results as a guideline, the RD53, ATLAS, and CMS collaborations can study what is possible given bandwidth and radiation hardness constraints and with full simulations of the detector readout, object reconstruction, and analysis emulation, what is needed to make the most of the high luminosity phase of the LHC.

\section{Acknowledgments}

This work was supported by the U.S.~Department of Energy, Office of Science under contract DE-AC02-05CH11231.

\clearpage

\appendix

\section{\label{sec:app} Downsampling Proof}

This section provides a constructive proof for the number of all possible functions to map $m$ bits of charge onto $n$ bits of charge, with $n\leq m$.  

\vspace{4mm}

\noindent \textbf{Definition}. Pick maps $f,g:\{0,...,2^{m}-1\}\rightarrow\{0,...,2^{n}-1\}$.  These two maps are \textit{redundant} if there exists a permutation $\sigma:\{0,...,2^{n}-1\}\rightarrow\{0,...,2^{n}-1\}$ not equal to the identity such that for all $i\in\{0,...,2^{m}-1\}$, $f(i)=\sigma(g(i))$.  

\vspace{4mm}

\noindent \textbf{Lemma}. Two surjective maps $f,g:\{0,...,2^{m}-1\}\rightarrow\{0,...,2^{n}-1\}$ that are not identical are also not redundant if they are monotonically increasing ($f(i+1)\geq f(i)$).

\vspace{4mm}

\begin{proof}
Suppose on the contrary that $f$ and $g$ are redundant.  Let $\sigma$ be a permutation such that $f(i)=\sigma(g(i))$ for all $i\in \{0,...,2^{m}-1\}$.  Pick any $i,j\in \{0,...,2^{m}-1\}$ with $j > i$.  By construction, $g(j)\geq g(i)$ and $f(j) \geq f(i)$.  Therefore, $\sigma$ is also a monotonically increasing map.  The only monotinically increasing bijective map is the identity which contradicts the fact that $f$ and $g$ are not identical. 
\end{proof}

\vspace{4mm}

\noindent \textbf{Lemma}. Consider the set of surjective $\mathcal{S}$ of monotonically increasing maps $f:\{0,...,2^{m}-1\}\rightarrow\{0,...,2^{n}-1\}$.  Any other surjective map $g:\{0,...,2^{m}-1\}\rightarrow\{0,...,2^{n}-1\}$ is redundant with a map in $\mathcal{S}$.

\vspace{4mm}

\begin{proof}
Consider a $g$ as in the statement of the lemma.  Construct a permutation $\sigma:\{0,...,2^{n}-1\}\rightarrow\{0,...,2^{n}-1\}$ using the following algorithm:

\begin{enumerate}
\item $\sigma(g(0))=0$.
\item For $0<i\leq2^{m}$, if $g(i-1)\neq g(i)$, set $\sigma(g(i))=\sigma(g(i-1))+1$.
\end{enumerate}

\noindent Note that since $g$ is surjective, $\sigma(g(2^{m}-1))=2^n-1$.

\end{proof}

\vspace{4mm}

\noindent \textbf{Theorem}. The total number of non-redundant surjective maps between the set of $m$ bit numbers and $n$ bit numbers is $\begin{pmatrix}2^{m-1}\cr 2^{n-1}\end{pmatrix}$.

\begin{proof}
There are $2^{n}$ possible choices in the range for each of the $2^m$ numbers in the domain.  Therefore, the total number of maps is $(2^n)^{2^m}$.  However, many of these maps are not surjective and there are many redundant maps.  By the lemmas, it is sufficient to count the number of monotonically increasing maps.  Let $f$ be one such map.  Due to surjectivity and monotonicity, $f(0)=0$ (cannot skip numbers).  For $i>0$, $f(i)=f(i-1)+0$ or $f(i)=f(i-1)+1$.  By the lemmas above, these two maps are not redundant.  Therefore, the set of possible maps can be enumerated by a length $2^m-1$ sequence of 0's and 1's with exactly $2^n$-1 1's.  There are $\begin{pmatrix}2^{m-1}\cr 2^{n-1}\end{pmatrix}$ such sequences.
\end{proof}

\clearpage

\bibliographystyle{elsarticle-num}
\bibliography{myrefs.bib}{}

\end{document}